\documentclass[a4paper,fleqn,usenatbib]{mnras}

\usepackage[T1]{fontenc}
\usepackage{ae,aecompl}

\usepackage{graphicx}	
\usepackage{amsmath}	
\usepackage{amssymb}	
\usepackage[table]{xcolor}

\usepackage{pdflscape}

\usepackage{booktabs}

\usepackage{gensymb}
\usepackage[normalem]{ulem}
\usepackage{newtxtext,newtxmath}

\title[Evaluating the jet/accretion coupling of Aql X-1]{Evaluating the jet/accretion coupling of Aql X-1: probing the contribution of accretion flow spectral components}

\author[Fijma et al.]{
S. Fijma,$^{1}$\thanks{E-mail: s.c.fijma@uva.nl}
J. van den Eijnden,$^{2,1}$
N. Degenaar,$^{1}$
T. D. Russell,$^{3,1}$
and J. C. A. Miller-Jones$^{4}$
\\
$^{1}$Anton Pannekoek Institute for Astronomy, University of Amsterdam, Science Park 904, 1098 XH, Amsterdam, the Netherlands\\
$^{2}$Astrophysics, Department of Physics, University of Oxford, Denys Wilkinson Building, Keble Road, Oxford OX1 3RH, UK\\
$^{3}$INAF, Istituto di Astrofisica Spaziale e Fisica Cosmica, Via U. La Malfa 153, I-90146 Palermo, Italy\\
$^{4}$International Centre for Radio Astronomy Research, Curtin University, GPO Box U1987, Perth, WA 6845, Australia\\
}

\date{Accepted XXX. Received YYY; in original form ZZZ}

\pubyear{2022}

\begin{document}
\label{firstpage}
\pagerange{\pageref{firstpage}--\pageref{lastpage}}
\maketitle

\begin{abstract}
The coupling between radio and X-ray luminosity is an important diagnostic tool to study the connection between the accretion inflow and jet outflow for low-mass X-ray binaries (LMXBs). The radio/X-ray correlation for individual neutron star (NS) LMXBs is scattered, whereas for individual black hole (BH) LMXBs a more consistent correlation is generally found. Furthermore, jet quenching is observed for both types of LMXBs, but it is unclear whether jets in NS-LMXBs quench as strongly as those in BH-LMXBs. 
While additional soft X-ray spectral components can be detected in NS-LMXB spectra due to the presence of the neutron star's surface, disentangling the individual X-ray spectral components has thus far not been considered when studying the radio/X-ray coupling. Here we present eleven epochs of \textit{Swift}/XRT observations matched with quasi-simultaneous archival radio observations of the 2009 November outburst of Aql X-1. We decompose the thermal and Comptonised spectral components in the \textit{Swift}/XRT spectra, with the aim of studying whether the presence of additional thermal emission affects the coupling of the radio/X-ray luminosity. 
We find that there is no evidence of a significant thermal contribution in \textit{Swift}/XRT spectra that could cause scatter in the radio/X-ray coupling. 
To explore the role of potential spectral degeneracies in the X-ray models and consider the improvements from including hard X-rays, we perform joint fits with quasi-simultaneous \textit{RXTE}/PCA spectra. Follow-up research using more sensitive, broad-band X-ray observations and densely-sampled near-simultaneous radio observations is required to study this in more detail.

\end{abstract}




\section{Introduction}
\label{sec:Introduction}



Coordinated, multi-wavelength observations of low mass X-ray binaries (LMXBs) allow one to probe the connection between the inflow and outflow for stellar-mass compact objects. 
LMXBs consist of a neutron star (NS) or a black hole (BH) and a < 1 M$_{\odot}$ companion star. In almost all LMXBs, material from the companion star is accreted onto the compact object through Roche-Lobe overflow. The material is transferred through the inner Lagrange point and forms an accretion disc around the compact object \citep{doi:10.1146/annurev.aa.09.090171.001151..RLOF}. Most systems show episodes of active accretion, which are proposed to be caused by instabilities in the disc \citep{1973A&A....24..337S..diskshakura, 1993ApJ...419..318C..hydrogen.ionisation.instability, 2000A&A...353..244H..hydrogen.ionisation.instability, 2001NewAR..45..449L..hydrogen.ionisation.instability},
alternated by quiescent periods. These episodes, or outbursts, are characterised by an increase in luminosity across the electromagnetic spectrum, driven by the increased mass accretion rate. The accretion flow emits photons from the infrared to the X-ray band, and jet outflows dominate in the radio band. 

For the X-ray emission specifically, one identifies several spectral components. Firstly, the disc emits a thermal or soft component, as material in the disc is heated up during outburst. Secondly, a population of hot electrons, also often referred to as the corona, close to the compact object Compton up-scatters this thermal emission from the disc, emitting a Comptonised component. Lastly, up-scattered photons interact with the accretion disc, resulting in reflection components \citep{2007A&ARv..15....1D..componentsoutburst, 2010LNP...794...17G..componentsoutburst}.
These are the components that are identified in the spectra of both BH and NS-LMXBs. 
For NS-LMXBs, one expects to find additional spectral components. This concerns emission from the neutron star surface, and/or from the ‘boundary layer’. This boundary layer is a result of the rotating material in the accretion disc reaching the neutron star surface and decelerating \citep{1998AstL...24..774S..boundarylayer,2008MNRAS.386.1038B..boundarylayer}. Both of these components can be identified in soft X-ray emission, typically below 3 keV. 

During outbursts of LMXBs, the ratio between the soft and hard spectral components varies over time, as the outburst progresses. Based on this ratio, combined with the timing properties of the X-ray binary, different accretion states can be distinguished. In the hard state, the Comptonised component is more prominent, while in the soft state, the thermal component is stronger \citep{2004MNRAS.355.1105F..HID,2007A&ARv..15....1D..componentsoutburst}. The physical processes which underpin these states, and the evolution which probe the transitions, are still being debated. The transition from the hard to the soft state is often interpreted as a change in the inner-disc radius, as well as a weakening of the hot electron cloud \citep{2007A&ARv..15....1D..componentsoutburst, 2010LNP...794...17G..componentsoutburst}. 

Moreover, for NS-LMXBs, there is also debate on the appropriate X-ray spectral models to describe the spectra of different subtypes and/or spectral states. Some studies find degeneracies between thermal model component(s) and the Comptonised component when fitting spectra from both the hard and soft state \citep{2007ApJ...667.1073L..Lin2007}. This currently presents different pictures of the structure and energetics of NS-LMXBs, further complicating the interpretation of the physical processes in these systems. Such issues may be exacerbated in X-ray observations of short exposure, limited bandpass and/or low flux, where such model degeneracies provide statistically equivalent descriptions of the data.

Classifying the spectral and timing properties of different LMXBs, allows one to track and study the transition between the hard, soft, and quiescent states. For BH-LMXBs, the evolution typically follows an increase in intensity from quiescence to the hard state, followed by a transition to the soft state through intermediate states at a somewhat constant intensity. Then the intensity decreases and the system transitions to the hard state and finally decays back to quiescence. In a hardness-intensity diagram \citep[like presented in][]{2004MNRAS.355.1105F..HID}, this follows a q-shape. For NS-LMXBs, the spectral behaviour is more complex. This behaviour can be identified both by the hardness-intensity diagram and the colour-colour diagram. For high accretion rates Z-shapes can be identified, and for lower accretion rates atoll-shapes can be identified in colour-colour diagrams. Comparing this to the properties of BH-LMXBs, the states identified for both these types of sources can be broadly interpreted as the equivalent of the hard, transitional/intermediate, and soft states respectively \citep{2010ApJ...719..201H..atoll}, and will be referred to as such in this work\footnote{See \cite{2015MNRAS.454.1371W} for a discussion on the nuance of this topic.}.
During an outburst of a NS-LMXB, it will transition through these states, and often show hysteresis in the hardness-intensity diagram \citep{2003MNRAS.338..189M..hysteresis}. 

When LMXB systems transition through the accretion states in the X-ray emission, one observes an evolution in the outflows seen for the radio emission as well. For BH-LMXBs, \cite{2004MNRAS.355.1105F..HID} propose a model describing the coupling between the accretion and ejection (radio/X-ray coupling), where compact jets are observed in the hard state \citep{2018MNRAS.478L.132G..hardstatejet}. When the source transitions to the soft state, the radio emission of the compact jet is observed to quench \citep{1971ApJ...164L...1H..BHdiscreteejecta, 1995Natur.374..703H....quenching, 1999ApJ...519L.165F...quenching,2004MNRAS.355.1105F..HID,2011ApJ...739L..19R..quenchingdiscussion} and discrete ejecta are observed \citep{2019ApJ...883..198R....discrete.ejecta, 2020NatAs...4..697B....discrete.ejecta, 2021MNRAS.504..444C....discrete.ejecta}. 
For NS-LMXBs the inflow/outflow coupling is less well understood \citep{2016LNP...905...65F..individualNScoupling}.
Compact jets are observed during the hard state. 
Strongly reduced radio emission is also observed in some atoll NS-LMXBs \citep[see e.g.][]{2011ATel.3198....1M..quench, 2010ApJ...716L.109M..MJ, 2017MNRAS.470.1871G..quench, 2018A&A...616A..23D..quench, 2021MNRAS.507.3899V}, although other such systems do not appear to show clear quenching of the compact jets, remaining radio-bright following the transition to the soft state \citep[see e.g.][]{2003MNRAS.342L..67M..quench, 2004MNRAS.351..186M..quench, 2021MNRAS.508L}. It is currently unclear why only a subset of NS-LMXBs appear to show radio quenching.

To derive this inflow/outflow coupling in the hard state, 
one may use the correlation measured between the X-ray and radio luminosity \citep{1998A&A...337..460H..correlation,2000A&A...359..251C..correlation, 2003A&A...400.1007C..correlation}. One finds that the radio/X-ray luminosity relation for LMXBs often follow a power law relation $L_{\mathrm{R}} \propto L_{\mathrm{X}}^{\beta}$ \citep{2002Sci...298..196C..lxlrpowerlaw, 2006MNRAS.366...79M..correlation} and that overall, when considering all sources for BH- and NS-LMXBs, both groups show approximately consistent power law slopes at the 2.5$\sigma$ level \citep{2018MNRAS.478L.132G..hardstatejet}. However, when considering individual sources in the BH- and NS-LMXB sample, the correlation is often more scattered \citep{2018MNRAS.478L.132G..hardstatejet} with sources showing significant deviations from the general population \citep{2021MNRAS.505L..58C} and sources showing different behaviours over different outbursts \citep{2020MNRAS.492.2858G}.
Moreover, additional challenges are introduced when analysing NS-LMXB sources. These systems are systematically more faint in radio for the radio/X-ray luminosity relation, by $\approx$ 22 for similar values for the X-ray luminosity \citep{2018MNRAS.478L.132G..hardstatejet}.
Therefore, it is more difficult to constrain the coupling between the radio and X-ray emission for individual sources, as most sources become undetectable at radio wavelengths for current observatories once their accretion rate decreases below $\sim 1$\% of the Eddington luminosity.

\subsection{Relation between jet and X-ray spectral components}
\label{sec:subsec_introduction}

For radio/X-ray coupling studies, the X-ray flux is typically obtained for the entire X-ray spectrum of each observation. Typically a simple absorbed power-law model is used to fit spectra of LMXB sources. However, more individual spectral components can be identified in the X-ray emission, which are shown to evolve during transitions in spectral states when LMXBs undergo outbursts. Currently, there is no decomposition of these different spectral components in analysing the radio/X-ray coupling for NS-LMXBs. The flux and luminosity for these individual spectral components could cause scatter for the coupling seen for individual systems, as these probe specific processes in the accretion flow morphology and spectral state of the system. 
Especially, NS-LMXBs show additional spectral components from the NS surface and/or the boundary layer as elaborated earlier. These components are not present in the spectra of BH-LMXBs, and might cause discrepancies when comparing the radio/X-ray coupling derived for both groups. 

A promising source for studying individual spectral components of a NS-LMXB system, is the source Aquila X-1 (Aql X-1) because it exhibits frequent outbursts. This system is one of the first Galactic X-ray sources to be discovered \citep{1967Sci...156..374F..aqlX-1galactic, 1973NPhS..245...37K..aqlX-1galactic}. It is a transient LMXB  \citep[orbital period $\approx$ 19 hours,][]{1991A&A...251L..11C..orbit} at a distance $d$ of around 4.5 kpc \citep{2008ApJS..179..360G...typeIbursts}.
It consists of a NS \citep{1981ApJ...247L..27K..AqlX-1NS} and a low-mass K-type companion star  \citep{1978ApJ...220L.131T..aqlX-1star, 1999A&A...347L..51C..aqlX-1star}. 
The source is classified as an atoll source \citep{2000ApJ...530..916R..atoll}, and goes into outburst roughly every year \citep{2013MNRAS.432.1695C..risedecay, 2017ApJ...848...13G..risedecay}. Therefore, it is used for many studies on the radio/X-ray coupling. 

\cite{2010ApJ...716L.109M..MJ} analysed the disc-jet coupling for the 2009 outburst of Aql X-1 by triggering an observation campaign with the Karl G. Jansky Very Large Array (VLA), the Very Long Baseline Array (VLBA) and the European Very Long Baseline Interferometry Network (EVN) at 4.8 and 8.4 GHz. The X-ray data were obtained with the \textit{RXTE} satellite, using the PCA (2-16 keV) and ASM (2-10 keV) count rates, as well as the count rates obtained by \textit{Swift}/BAT (15-150 keV). 
\cite{2010ApJ...716L.109M..MJ} found that the radio emission is consistent with being triggered during the state transition, and confirms radio quenching at X-ray luminosities of around 10 percent of the Eddington luminosity (for a 1.4 M$_\odot$ NS accretor this is around $L_{\mathrm{X}} \approx 10^{37} (d/ 5 \text{kpc})^2$) erg/s at a distance $d$). 
The radio emission shows interesting behaviour during this outburst, this is discussed in more detail in Section 
\ref{sec:discussion}. However, the analysis by \cite{2010ApJ...716L.109M..MJ} used X-ray count rates and colors, measured above 2 keV, to study the X-ray/radio coupling. A full X-ray spectral analysis focusing on this outburst, including especially the soft X-rays, has not yet been performed. 
Interestingly, \textit{Swift}/XRT observed this outburst and measured the X-ray spectra in the soft X-ray band (0.5-10 keV). 
These observations were performed in a similar time span to the radio observations, and can therefore be used to perform a radio/X-ray luminosity study.

In this paper, we use archival \textit{Swift}/XRT data of the 2009 outburst of Aql X-1 to perform a detailed analysis on the X-ray spectral components. We obtain the X-ray emission and luminosity from these separate components after performing a careful consideration of the presence of the thermal component. We then use these measurements to constrain the thermal flux and its evolution during the outburst. 
We further evaluate the implications of model degeneracies by comparing our results with joint fits including data from \textit{RXTE}.
Furthermore, we use archival VLA and VLBA data from \cite{2010ApJ...716L.109M..MJ} to analyse the radio/X-ray connection for this outburst. Using the X-ray flux obtained for the different spectral components, we evaluate the individual radio/X-ray coupling obtained for each, specifically investigating whether the presence of additional thermal emission in this NS-LMXB causes scatter in the X-ray/radio luminosity coupling. 
In Section \ref{sec:datamethod} we present the observations and data analysis, in Section \ref{sec:results} the results, and we will discuss this work in Section \ref{sec:discussion}.

\section{Observations and Analysis} 
\label{sec:datamethod}

The Proportional Counter Array \citep[PCA;][]{1996SPIE.2808...59J..RXTE} onboard the \textit{RXTE} satellite observed rising flux from Aql X-1 on November 1st, 2009 during an ongoing Galaxy bulge monitoring program \citep{2000arxt.confE...7M}. This rise in flux triggered multiple observing campaigns to monitor the outburst, among which the \textit{Swift} telescope \citep[\textit{Swift};][]{2004ApJ...611.1005G..swifttelescope}, the VLA and the VLBA.

The VLA and VLBA monitoring is presented by \cite{2010ApJ...716L.109M..MJ}. The outburst was observed at an observing frequency of 8.4 GHz for both observatories. More information on the calibration and reduction of the data can be found in the original paper. In this work we will use the VLA and VLBA radio flux densities reported by \cite{2010ApJ...716L.109M..MJ} close in time to the \textit{Swift} observations discussed below. We will also discuss the selection method later in this Section. The dates and radio flux densities for these observations are listed in Table \ref{tab:observations}.

Using \textit{Swift}’s X-Ray Telescope (XRT) the outburst was observed within an energy range from 0.5 to 10 keV. 
Sixteen observations were taken in total. The first fifteen observations were obtained in Windowed Timing mode (WT) and the last observation was obtained in both WT and Photon Counting (PC) mode. 
We obtain the data from the \textit{Swift}-XRT data products generator\footnote{\url{https://www.swift.ac.uk/user_objects/index.php}} \citep{2007A&A...469..379E..xspec, 2009MNRAS.397.1177E..xspec}.
We then fit the spectra using \texttt{XSPEC} \citep[v12.11.0;][]{1996ASPC..101...xspec.ref}. We set the abundances and cross-sections based on \cite{2000ApJ...542..Wilms} and \cite{1996ApJ...465..Verner}, respectively. To determine the goodness-of-fit and to estimate the best-fit spectral parameters we used the $\chi^{2}$ statistic. 
We also obtain the \textit{Swift}/BAT (15-50 keV) light curve \citep{2013ApJS..209...14K..BAT} to illustrate the evolution of the 2009 outburst and to study the hard X-ray colour, we discuss this in Section \ref{sec:results}.
Finally, we include data from the \textit{RXTE}/PCA instrument to assess potential model degeneracies. For this purpose, we download the standard-2 data products, including source and background spectra, for the \textit{RXTE} observations presented in \citep{2010ApJ...716L.109M..MJ} and taken during the outburst phase observed by \textit{Swift}. We apply the above approach to the PCA spectral analysis, and additionally include systematic errors of $0.8\%$ for PCA channels 0-39 and $2\%$ for channels 40-128 \citep{2004A&A...427..975K..RXTE..systematic.errors, 1996SPIE.2808...59J..RXTE}. We also use the associated background spectra to determine the energy range to fit the source spectrum, corresponding to 2.6-23 keV.

Our research aims to distinguish the contribution of a black body component, probing the thermal emission of the accretion disc, neutron star surface and/or the boundary layer, and of the component induced by Comptonisation probing the corona. Therefore we used corresponding components in the model (\texttt{diskbb} and \texttt{nthcomp}, respectively), as well as a component (\texttt{tbabs}) to account for the absorption by the interstellar medium. The complete model is \texttt{tbabs*(diskbb+nthcomp)} in \texttt{XSPEC} syntax, and will be referred to as the composite model. We used an absorbed power-law model as well, \texttt{tbabs*nthcomp}, to test the significance of using a model containing a black body component. This model will be referred to as the Comptonised model. We determine the significance of the presence of a thermal component using \texttt{ftest} to calculate the F-statistic and its probability. Furthermore, we used \texttt{cflux} to calculate the flux of the total model, the \texttt{diskbb} component and the \texttt{nthcomp} component. 

\begin{figure}
 \includegraphics[width=\columnwidth]{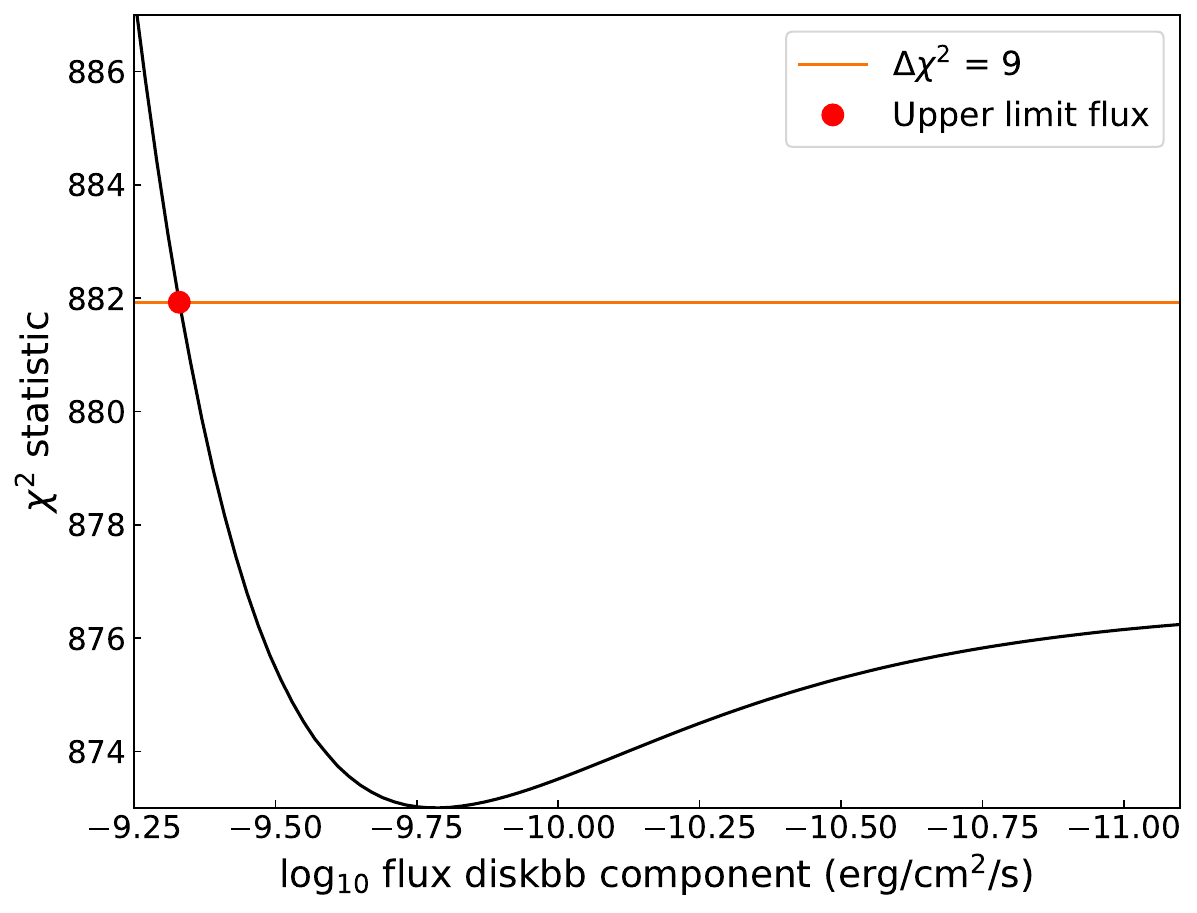}
 \caption{An example of using the \texttt{steppar} command to determine the 3$\sigma$ upper limit of the flux for the \texttt{diskbb} component for observation X7. We use \texttt{steppar} to manually vary the parameter around the best fitting value for the flux, i.e. where the $\chi^2$ statistic has the lowest value. We determine the 3$\sigma$ upper limit value where the $\chi^2$ statistic increased by $\Delta \chi^2$ = 9.  The parameter value of the log$_{10}$ flux is shown in black, the line where $\Delta \chi^2$ = 9 as opposed to the minimum value for the $\chi^2$ statistic is shown in orange, and the obtained value for the 3$\sigma$ upper limit is indicated with a red dot.}
 \label{fig:steppar}
\end{figure}

We only calculate the flux of the \texttt{diskbb} for observations where the composite model was statistically preferred over the Comptonised model. If the thermal component is not significantly present in the spectrum, we instead aimed to determine an upper limit of its flux to compare with the radio luminosity at that time. For this purpose, we used \texttt{steppar} to manually vary the flux parameter in the \texttt{tbabs*(cflux*bbody + powerlaw)} model (i.e. where the flux is only measured over the thermal component). The \texttt{steppar} task gradually increases the flux, at each step fixing it and subsequently fitting all other parameters, before saving the new fit statistic. Repeating this procedure for increasing fluxes, we determined the thermal flux where the fit statistic increased (i.e. worsened) by $\Delta \chi^2$ = 9, corresponding to 3$\sigma$ when varying a single parameter. We show an example in Figure \ref{fig:steppar}, using observation X7 (see Table \ref{tab:observations}). In this Figure, the improvement in fit statistic at the best-fit \texttt{diskbb} flux, compared to the plateau in $\chi^2$ at a flux of zero (i.e. towards the right), is less than three sigma.

We note that the flux in a black body spectrum scales with both normalisation (linearly) and temperature (to the fourth power), creating a parameter degeneracy; therefore, using the \texttt{steppar} approach on either of those parameters individually does not directly yield a flux upper limit. With the \texttt{steppar} approach, the best fitting combination of temperature and normalisation is found for the considered total flux in each step.

In order to determine the absorption column $n_{\mathrm{H}}$, we fit the Comptonised model to all spectra simultaneously with the \texttt{tbabs} parameter tied. The best fit was found for a value of $n_{\mathrm{H}} = 0.355 \times 10^{22}$ cm$^{-2}$ for the equivalent hydrogen column. As this is consistent with the value for $n_{\mathrm{H}}$ found in other studies \citep[see e.g.][]{2012PASJ...64...72S...nH}, we use this value for our analysis. 

When fitting the Comptonised and composite model to the \textit{Swift}/XRT spectra, we froze the value for $n_{\mathrm{H}}$ and fixed this to $0.355 \times 10^{22}$ cm$^{-2}$. Furthermore, we linked the parameter corresponding to the temperature at the inner disc radius of the \texttt{diskbb} component, the parameter corresponding to the seed photon temperature of the \texttt{nthcomp} component ($kT_{\text{bb}}$). We link these parameters as they are believed to probe the same emission process, assuming that the seed-photons seen by the corona have a thermal origin. For both models, the \texttt{nthcomp} parameters \texttt{inp\_type} and \texttt{redshift} are set to 0. \texttt{inp\_type} is set to 0 assuming black body seed photons for the model, and the \texttt{redshift} is set to 0 based on the distance to Aql X-1.

We fitted the composite model to all WT mode spectra to obtain the parameter values. Photons between 0.7-10 keV were used for spectral fitting, as at lower energies, variable calibration residuals might be present which could affect our conclusions regarding the presence of a soft thermal component \footnote{See e.g. \url{https://www.swift.ac.uk/analysis/xrt/digest_cal.php} for details.}.

We also checked if thermonuclear (Type I) X-ray bursts are present in the light curves of the \textit{Swift}/XRT observations. As the burst X-ray spectrum is generally consistent with a black body of $T_{\text{bb}} \approx$ 2-3 keV \citep{2008ApJS..179..360G...typeIbursts}, this could affect our conclusions regarding the presence of a soft thermal component as well. Therefore, we created a light curve of each of the observations using the \textit{Swift}-XRT data products generator. We use time-bins of 1 second to create light curves of the WT observations, as Type-I bursts typically exhibit rise times between around one and ten seconds, and can last between tens to hundreds of seconds. 

Lastly, we calculate the X-ray luminosity of each observation from the obtained flux using $L_{\mathrm{X}} = 4 \pi d^{2} F_{\mathrm{X}}$, with $d$ being the distance to Aql X-1. We used the distance of 4.5 kpc as determined by \cite{2008ApJS..179..360G...typeIbursts}.
To obtain the radio luminosity for the VLA and VLBA observations, we used $L_{\mathrm{R}} = 4 \pi d^{2} \nu S_{\nu}$, where $S_{\nu}$ is the radio flux density and $\nu$ is the frequency at which the flux has been observed. 
To be able to compare the X-ray and radio luminosity, we matched the X-ray and radio observations if they were measured $\la 1$ day from each other; we will discuss the effects of non-simultaneity explicitly where relevant.

\section{Results}
\label{sec:results}

\subsection{X-ray and Radio lightcurves}
\label{sec:results_lightcurves}

\begin{figure}
 \includegraphics[width=\columnwidth]{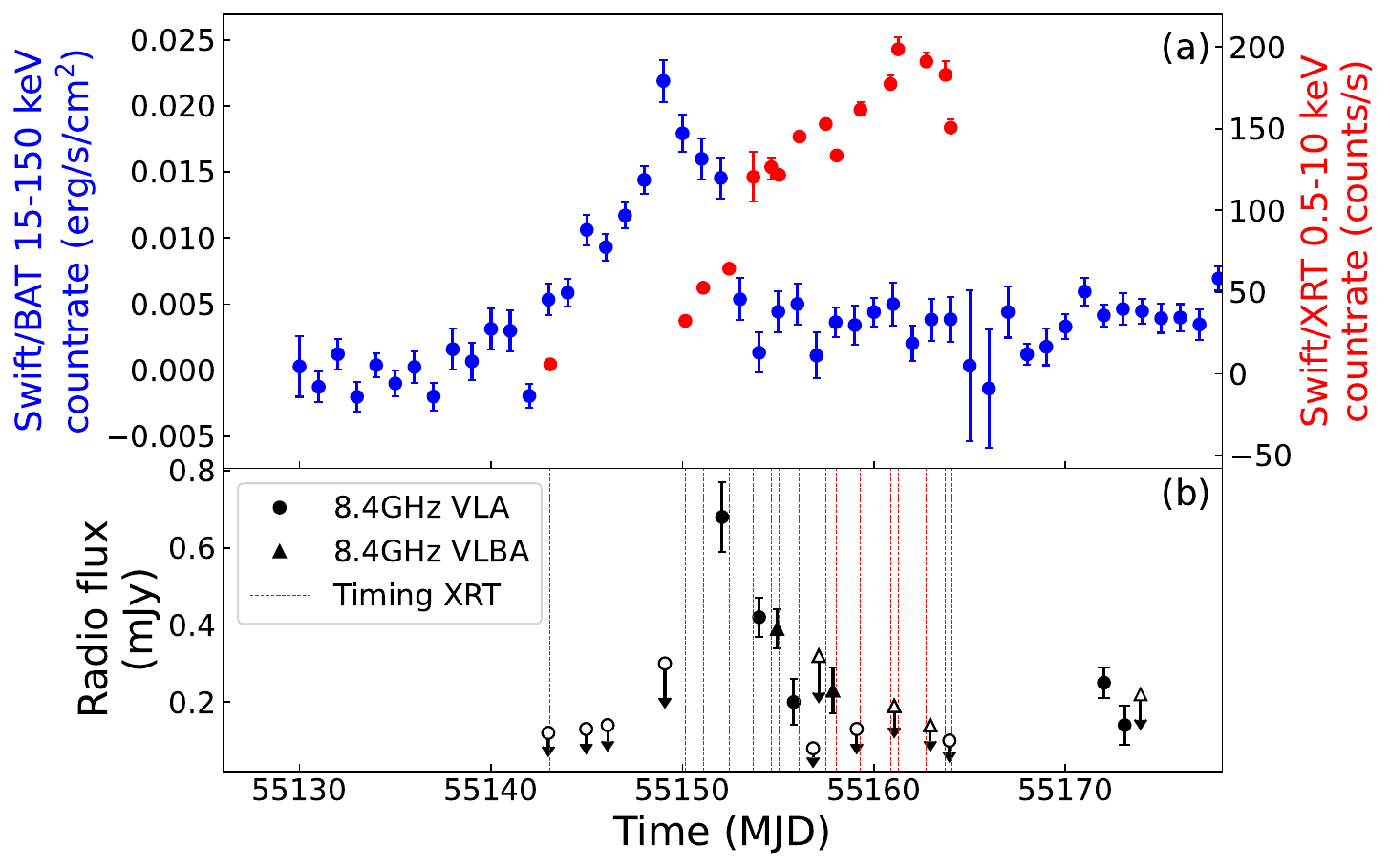}
 \caption{The radio and X-ray light curves of the 2009 outburst of Aql X-1. (a) The observed \textit{Swift}/XRT and BAT count rates, indicated with red and blue markers respectively. (b) The radio flux densities as first recorded by \citet{2010ApJ...716L.109M..MJ}. VLA 8.4 GHz and VLBA 8.4 GHz observations are indicated by circles and triangles respectively. Filled markers indicate detections and the corresponding open markers the (3$\sigma$) upper limits. The red dotted lines indicate the timing of the XRT observations.}
 \label{fig:lightcurves}
\end{figure}

In Figure \ref{fig:lightcurves}, we show the \textit{Swift}/XRT and BAT X-ray light curves, as well as the VLA and VLBA radio light curves first recorded by \cite{2010ApJ...716L.109M..MJ}. The hard X-ray flux peaks first during the outburst as typically observed in Aql X-1 \citep{2003ApJ...589L..33Y}, around MJD 55150. The hard X-ray flux then decays as the soft X-ray flux rises at around MJD 55153. As noted by \cite{2010ApJ...716L.109M..MJ}, radio emission is first detected by the VLA during the softening of the X-ray spectrum. The radio emission then decays. \cite{2010ApJ...716L.109M..MJ} note the source transitions from the hard to the intermediate state around MJD 55151, and from the intermediate to the soft state around MJD 55155, based on the \textit{RXTE} observations. 
Finally, the soft X-ray emission peaks around MJD 55161. \cite{2010ApJ...716L.109M..MJ} describe a second increase in radio emission once the spectrum hardens again around MJD 55179.
However, the additional radio data and the BAT data after MJD 55164 are excluded from this analysis, as there are no available XRT observations to match the radio emission from the second rise due to Sun constraints. 

The aim of this study is to determine the radio/X-ray coupling during this outburst for individual X-ray spectral components. We show the corresponding pairing of observations in Table \ref{tab:observations}, where observations listed with $X$ and $R$ indicate \textit{Swift} X-ray and \text{VL(B)A} radio observations, respectively. R13 is the only radio observation which was measured within one day of two X-ray observations, but could not be distinctly matched to either: the separation from both XRT observations is approximately equal. 
Moreover, we find that matching R13 to either X12 or X13 does not significantly impact our results or conclusions in the rest of this work.
Therefore, observation R13 is excluded from the analysis. 
As a second test of the effect of sampling, we decided to combine \textit{Swift}/XRT spectra to better match the VLA observations in time. We have combined observations X3 and X4 to match R5, X5, X6 and X7 to match R6, X8 and X9 to match R9, and X12, X13, X14 and X15 to match R15. However, this provided no additional insights for this work.

\begin{table*}
\begin{tabular}{clcccc|c|lcccc}
\hline
 & X-ray & Date & MJD & Duration obs. & ObsID & Time difference & Radio & Date & MJD & Duration obs. & Instrument \\ 
 & obs. & (2009) & (day) & (ks) &  & X-R (hour) & obs. & (2009) & (day) & (hour) &  \\ \hline
\cellcolor[RGB]{255,232,232} & X1 & Nov 8 & 55143.09 & 2.52 & 30796036 & 2.43 & R1 & Nov 7 & 55142.99 & 0.4 & VLA \\ 
 & - & - & - & - & - & - & R2$^{a}$ & Nov 9 & 55144.97 & 0.35 & VLA \\ 
 & - & - & - & - & - & - & R3$^{a}$ & Nov 11 & 55146.09 & 0.34 & VLA \\ 
\cellcolor[RGB]{255,217,213} & X2 & Nov 15 & 55150.15 & 1.00 & 30796231 & 25.65 & R4 & Nov 14 & 55149.08 & 0.41 & VLA \\ 
 & X3$^{a}$ & Nov 16 & 55151.09 & 1.00 & 30796232 & - & - & - & - & - & - \\ 
\cellcolor[RGB]{255,201,192} & X4 & Nov 17 & 55152.44 & 1.56 & 30796234 & 8.82 & R5 & Nov 17 & 55152.07 & 0.39 & VLA \\ 
\cellcolor[RGB]{253,183,169} & X5 & Nov 18 & 55153.70 & 1.58 & 30796235 & -7.17 & R6 & Nov 19 & 55154 & 0.82 & VLA \\ 
 & X6$^{a}$ & Nov 19 & 55154.64 & 1.52 & 30796236 & - & - & - & - & - & - \\ 
\cellcolor[RGB]{249,162,144} & X7 & Nov 20 & 55155.04 & 1.01 & 30796237 & 2.51 & R7 & Nov 19 & 55154.94 & 4.95 & VLBA \\ 
\cellcolor[RGB]{245,142,120} & X8 & Nov 21 & 55156.11 & 1.38 & 30796238 & 7.25 & R8 & Nov 20 & 55155.81 & 0.81 & VLA \\ 
 & - & - & - & - & - & - & R9$^{a}$ & Nov 21 & 55156.82 & 1.07 & VLA \\ 
\cellcolor[RGB]{239,121,96} & X9 & Nov 22 & 55157.48 & 0.85 & 30796239 & 8.66 & R10 & Nov 22 & 55157.12 & 1.75 & VLBA \\ 
\cellcolor[RGB]{232,99,73} & X10 & Nov 23 & 55158.05 & 1.32 & 30796240 & 4.62 & R11 & Nov 22 & 55157.86 & 2.26 & VLBA \\ 
\cellcolor[RGB]{224,78,52} & X11 & Nov 24 & 55159.29 & 1.33 & 30796241 & 4.66 & R12 & Nov 24 & 55159.1 & 0.91 & VLA \\ 
 & X12$^{a}$ & Nov 25 & 55160.86 & 1.19 & 30796242 & - & - & - & - & - & - \\ 
 & - & - & - & - & - & - & R13$^{a}$ & Nov 26 & 55161.06 & 3.84 & VLBA \\ 
 & X13$^{a}$ & Nov 26 & 55161.26 & 2.28 & 30796243 & - & - & - & - & - & - \\ 
\cellcolor[RGB]{216,54,30} & X14 & Nov 27 & 55162.74 & 1.54 & 30796244 & -4.83 & R14 & Nov 27 & 55162.94 & 4.92 & VLBA \\ 
 & X15$^{a}$ & Nov 28 & 55163.74 & 1.27 & 30796245 & - & - & - & - & - & - \\ 
\cellcolor[RGB]{207,14,4} & X16 & Nov 29 & 55164.01 & 1.05 & 30796246 & 1.67 & R15 & Nov 28 & 55163.94 & 0.89 & VLA \\ 
\hline
\end{tabular}
\caption{Details of the \textit{Swift}/XRT observations and the corresponding quasi-simultaneous VLA and VLBA 8.4-GHz radio detections and upper limits ($3\sigma$) of Aql X-1. We obtain the archival radio data from \citet{2010ApJ...716L.109M..MJ}. The MJD for the radio observations are provided for the middle of the observations, and the duration is calculated using the start and stop time for the scans of Aql X-1. The MJD for the X-ray observations are provided for the middle of the observations as well. All X-ray spectra are analysed as elaborated in Section \ref{sec:datamethod}. The X-ray and radio observations measured $\lesssim 1$ day apart are matched for the analysis of the radio/X-ray coupling. These data are indicated with red sequential colours from light to dark to indicate the evolution in time in later Figures. $^{a}$~These X-ray observations could not be matched to a Radio observation and are therefore excluded from this analysis.}
\label{tab:observations}
\end{table*}

\subsection{X-ray spectral model fitting} 
\label{sec:results_model}

To calculate the X-ray flux, we fit the \textit{Swift}/XRT spectra using the composite model as elaborated in Section \ref{sec:datamethod}. 
We aim to determine if adding a black body model component is statistically significant for the XRT spectra, by comparing the models and calculating the F-statistic and its probability. Typically \textit{Swift}/XRT spectra of NS-LMXBs can be fitted adequately with a power law model, especially at low luminosities \citep{2015MNRAS.454.1371W}. The $\chi^{2}$-statistic for both models and the resulting \textit{p}-value per observation are shown in Table \ref{tab:chisqr}. For the parameter values obtained for the composite model, see Figure \ref{fig:parameters}.

\begin{table}
\begin{tabular}{@{}lcccc@{}}
\toprule
Obs       & $\chi^{2}$ dual & $\chi^{2}$ single & p-value  & Significance level              \\
          & model (d.o.f.)  & model (d.o.f.)    &          & p value ($\sigma$) \\ \midrule
X1        & 933.33 (731)    & 938.59 (732)      & 4.27E-02 & 1.72                            \\
X2        & 845.07 (807)    & 853.86 (808)      & 3.87E-03 & 2.66                            \\
X3$^{a}$  & 514.58 (567)    & 517.08 (568)      & 9.75E-02 & 1.30 \\                            
X4        & 932.67 (872)    & 999.35 (873)      & 8.65E-15 & 7.67                            \\
X5        & 935.52 (859)    & 935.56 (860)      & 8.48E-01 & 1.03                            \\
X6$^{a}$  & 1088.66 (896)   & 1088.66 (897)     & 1        & -                               \\
X7        & 873.01 (881)    & 876.6 (882)       & 5.73E-02 & 1.58                            \\
X8        & 872.95 (845)    & 876.27 (846)      & 7.34E-02 & 1.45                            \\
X9        & 844.33 (806)    & 849.91 (807)      & 2.13E-02 & 2.03                            \\
X10       & 1386.74 (907)   & 1386.86 (908)     & 7.79E-01 & 0.77                            \\
X11       & 898.53 (856)    & 898.67 (857)      & 7.15E-01 & 0.57                            \\
X12$^{a}$ & 825.33 (862)    & 825.33 (863)      & 1        & -                               \\
X13$^{a}$ & 930.36 (896)    & 950.14 (897)      & 1.42E-05 & 4.19                            \\
X14       & 904.7 (889)     & 912.83 (890)      & 4.81E-03 & 2.59                            \\
X15$^{a}$ & 924.71 (842)    & 933.33 (843)      & 5.20E-03 & 2.56                            \\
X16       & 845.12 (871)    & 845.12 (872)      & 1        & -                               \\ \bottomrule
\end{tabular}%
\caption{Details of the X-ray spectral fits for the composite (\texttt{tbabs*(nthcomp+diskbb)}) and Comptonised model (\texttt{tbabs*nthcomp}) as elaborated in Section \ref{sec:datamethod}. The probability of the F-statistic, or \textit{p} value, is obtained by comparing the $\chi^{2}$ values and degrees of freedom (d.o.f.) obtained for both models for each observation using \texttt{ftest} in \texttt{XSPEC}. The significance level is obtained using a standard normal distribution ($\mu$=0, $\sigma$=1) for the corresponding probability with a one-tailed test. If the \textit{p} value exceeds a $2.5\sigma$ confidence level, we consider the contribution of the \texttt{diskbb} component to the model to be significant. See Section \ref{sec:results} for more details. 
\\\hspace{\textwidth} $^{a}$  These X-ray observations could not be matched to a Radio observation and are therefore excluded from this analysis.}
\label{tab:chisqr}
\end{table}

\begin{figure}
 \includegraphics[width=\columnwidth]{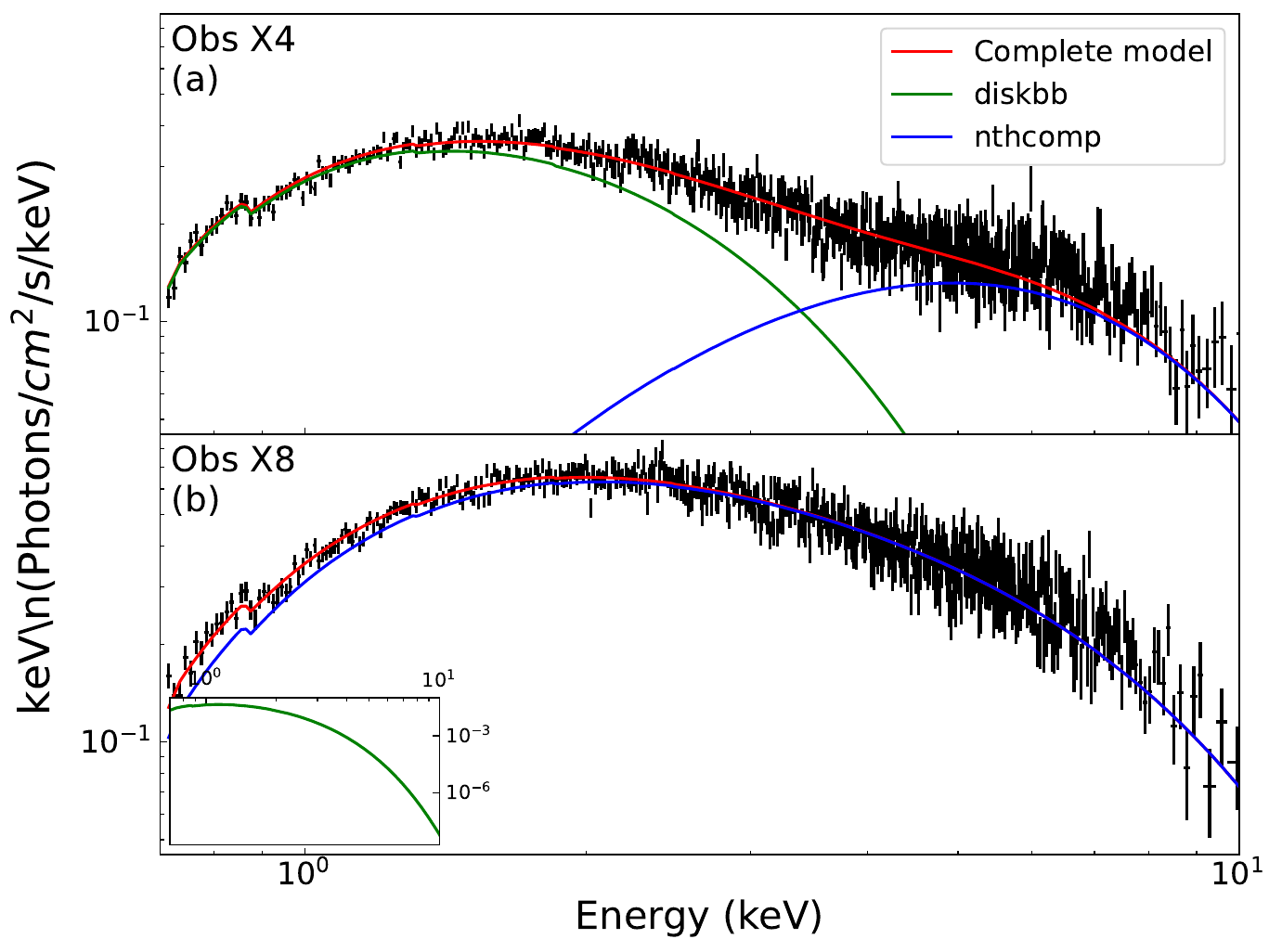}
 \caption{The spectra of observations X4 (Top panel a) and X8 (bottom panel b) plotted with the composite model and the individual model components. The data is shown in black, the complete model in red, and the \texttt{diskbb} and \texttt{nthcomp} model in green and blue, respectively. The data has been visually rebinned (at 5$\sigma$) for clarification. For observation X4 adding the black body component corresponds to a confidence level of 7.67$\sigma$ and is considered significant. For X8 adding the component corresponds to a confidence level of 1.58$\sigma$ and is not considered significant. 
 For observations X8 the \texttt{diskbb} component is so weak compared to the \texttt{nthcomp} component, that we show the \texttt{diskbb} component in the inset for visual purposes, with the scale shown on the right axis. }
 \label{fig:spectra}
\end{figure}

Initially we consider spectra to have a detectable black body component if the probability of adding the \texttt{diskbb} model component (i.e. using the composite model instead of the Comptonised model) results in a confidence level exceeding $3\sigma$. This requirement is met for observations X4 and X13. However, for observations X2, X14 and X15, this confidence level of adding the thermal component is almost obtained and exceeds $2.5\sigma$, while it is not as significant ($\lesssim 2 \sigma$) for the remaining observations. Therefore, for this analysis, we consider X2, X14 and X15 as having a detectable black body component as well. 
We will further discuss this assumption, and the effect of a broader energy band on these significances, in Section \ref{sec:discussion}.

From our results, it appears that the thermal component intermittently contributes to the model when fitting the spectra. 
We note a significant contribution for observations X2 and X4 when the hard X-ray flux decays and the soft X-ray flux increases, around MJD 55150 to 55153. We later see a significant contribution for observations X13, X14 and X15, where the soft X-ray flux peaks for the recorded XRT data around MJD 55161 to 55164. 

It is important to consider that this could be the result of degeneracies between the thermal and Comptonised model component. In previous studies using \textit{RXTE}/PCA spectra, the soft state spectra could be fitted using a cool thermal component and a dominant Comptonised component \citep{2002MNRAS.337.1373G..Done2002,2007ApJ...667.1073L..Lin2007}. In this case, it is difficult to constrain the temperature and normalisation of the thermal component, and sometimes the thermal component is not required to include in the model.
Because of the degeneracies found between the thermal and Comptonised model component for NS-LMXBs, this might not be an accurate model for the spectra in the soft state, also considering other studies finding a strong thermal component in the soft state \citep[e.g.][]{2011MNRAS.416..637R..raichur2011}. 

To test whether these degeneracies play a role, we also attempted to first fix parameters to fit a strong thermal component in all observations, and later allowed them to vary, to determine if this changed our results. However, we found that this provided similar results as our earlier fits, which only requires a thermal component for observations X2, X4, X13, X14 and X15. In other words, the lack of detectable thermal component does not arise from initial guesses of parameters and the presence of local minima in the multi-dimensional parameter space.

We can further review the shape of the X-ray spectra.
We show an example in Figure \ref{fig:spectra} comparing the spectra of observations X4 and X8, where X4 has a detectable thermal component and X8 does not.
Here, there appears to be a distinction in shape between the two spectra, which could be attributed to the contribution of a strong thermal component in X4. However, since \textit{Swift/XRT} spectra only cover energies up until 10 keV, this might be biased by the limited energy range over which the fits are performed. 
We will discuss this in Section \ref{sec:discussion}.

\begin{figure}
 \includegraphics[width=\columnwidth]{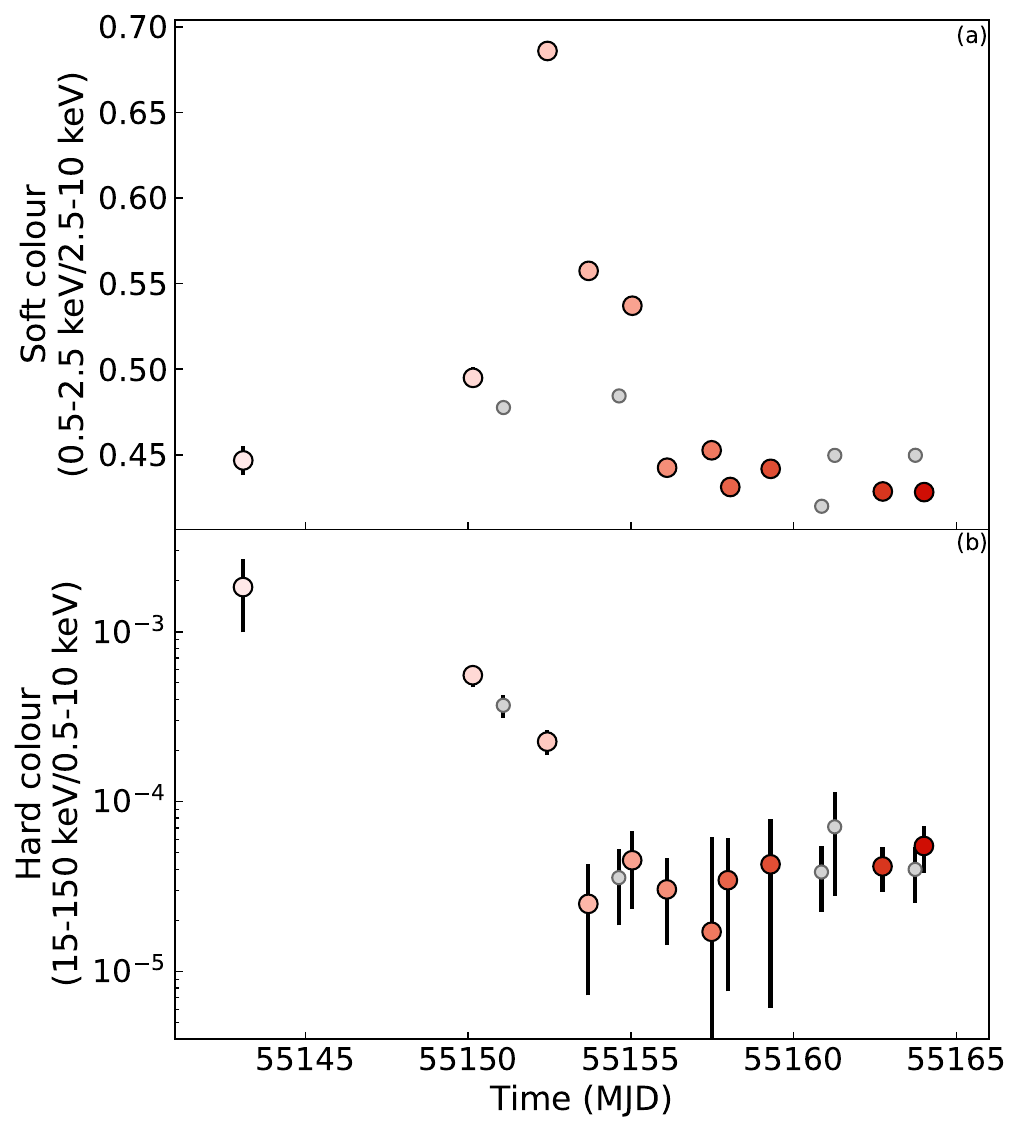}
 \caption{The evolution of \textit{Swift} X-ray colours throughout the 2009 outburst of Aql X-1. The evolution in time is indicated with red sequential colours from light to dark. The large coloured points indicate the X-ray observations that could be linked to quasi-simultaneous radio observations. Small grey points indicate X-ray observations that could not be linked. (a) The soft X-ray colour, we define this as the ratio between the \textit{Swift}/XRT count rates in the soft and hard bands. (b) The hard X-ray colour, we define this as the ratio between the \textit{Swift}/XRT countrate in WT mode and the \textit{Swift}/BAT orbital countrate.} 
 \label{fig:HR}
\end{figure}

\begin{figure}
 \includegraphics[width=\columnwidth]{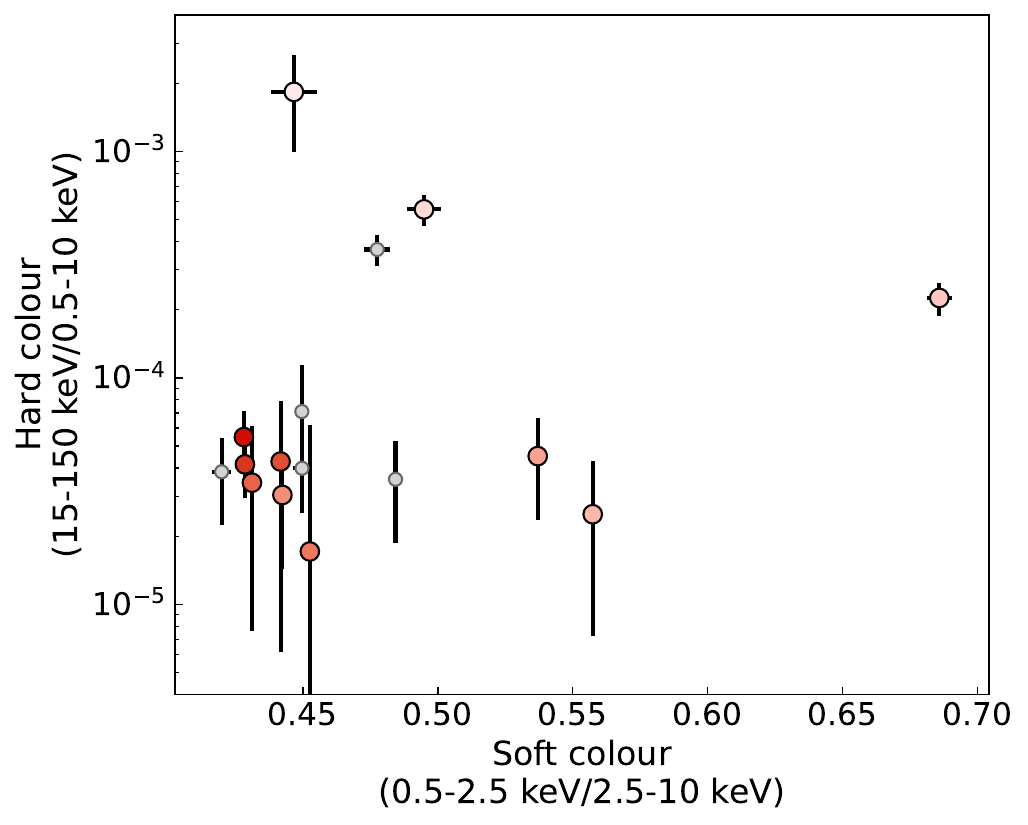}
 \caption{ The colour-colour diagram for the 2009 outburst of Aql X-1. The colour coding of the data points and the definition of soft/hard colour used here is the same as in Figure \ref{fig:HR}.} 
 \label{fig:colorcolor}
\end{figure}

We can also study the evolution in the X-ray colours, to evaluate whether the seemingly intermittent contribution of the thermal component could be related to spectral state transition(s). 
We define the soft X-ray colour as the ratio between the \textit{Swift}/XRT countrate in the soft (0.5-2.5 keV) and hard (2.5-10 keV) bands. Secondly, we define the hard X-ray colour as the ratio between the \textit{Swift}/XRT countrate in WT mode (0.5-10 keV) and the \textit{Swift}/BAT countrate (15-150 keV). We show the evolution of the individual colours in Figure \ref{fig:HR}, and the colour-colour diagram is shown in Figure \ref{fig:colorcolor}. We note that the soft colour initially increases until MJD 55153, after which it decreases again. The soft colour has the highest value for observation X4, where we note the most significant contribution of the black body component. As the soft colour roughly reflects the shape of the spectra measured with XRT between 0.5-10 keV, this result is expected. However, the remaining observations with a significant black body component show lower values for the soft colour. For the hard colour we initially see a decrease, and around MJD 55154 the hard colour levels out. This is reflected in the evolution of the BAT and XRT light curves shown in Figure \ref{fig:lightcurves} as well. 
Based on our results for the BAT and XRT lightcurves, as well as the results for the soft and hard X-ray colour, we conclude that Aql X-1 underwent a transition from a hard to a soft state before MJD 55153.7, and was in the soft state from MJD 55153.7 onward. As mentioned in Section \ref{sec:results_lightcurves}, the XRT observations stopped after MJD 55164, so we missed the transition back to the hard state as reported by \cite{2010ApJ...716L.109M..MJ}.

We note that X2 and X4 both show a significant black body component in their spectra and occur during the softening of the soft and hard colour. However, this is not apparent for X13, X14 and X15, which occur after the X-ray colours appear levelled. Furthermore, no significant black body component has been identified for observation X3, which occurs during the spectral softening as well. Therefore, we cannot determine a connection between the transitions of the spectral state and the presence of a significant black body component in the XRT spectra. We return to this point in the Section \ref{sec:discussion}.

As elaborated in Section \ref{sec:datamethod}, we checked the light curves of the individual X-ray observations for Type-I bursts. We find a Type-I burst in one of the observations, namely in observation X3. However,
the thermal component is not detected in this observation with the burst included, and removing the burst does not affect the parameter values for the composite and Comptonised model significantly. For our analysis we have removed the burst in this observation using \texttt{XSELECT}
and present the results accordingly. 

\begin{figure}
 \includegraphics[width=\columnwidth]{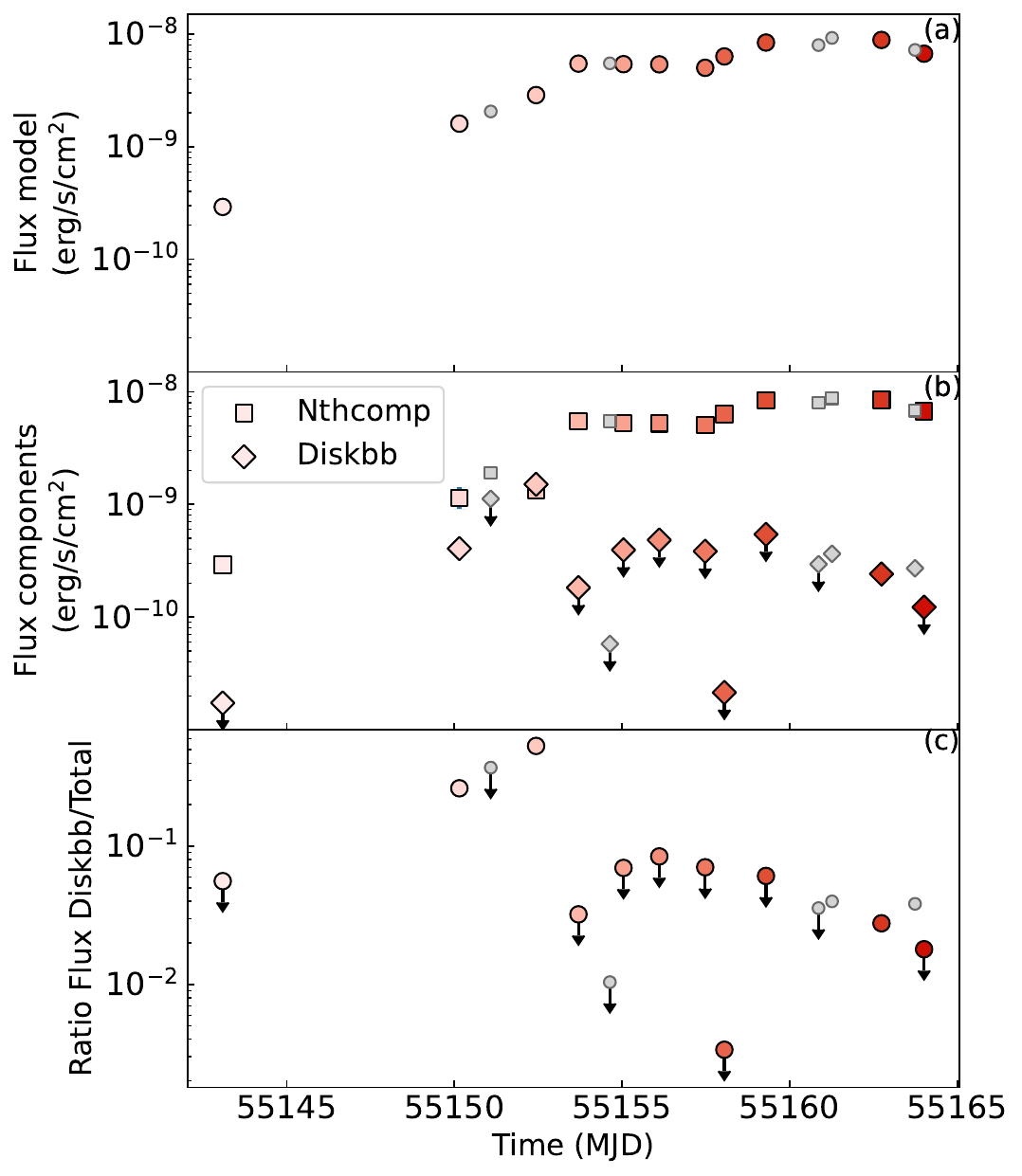}
 \caption{The X-ray flux values obtained for the composite model and the separate model components between 0.7-10 keV. The colour coding of the data points used here is the same as in Figure \ref{fig:HR}. (a) The X-ray flux calculated for the complete composite model. (b) The X-ray flux for the \texttt{diskbb} and \texttt{nthcomp} model components from 0.7-10 keV. The \texttt{diskbb} and \texttt{nthcomp} component are indicated using diamonds and squares, respectively. (c) The ratio between the X-ray flux for the \texttt{diskbb} component and the total flux.} 
 \label{fig:flux}
\end{figure}

For the observations where the composite model does not fit significantly better, we determine the $3\sigma$ upper limit of the black body flux as explained in Section \ref{sec:datamethod}, by determining the flux where the fit statistic increased (i.e. worsened) by $\Delta \chi^{2}$ = 9. 
We show the total flux in the composite model, the flux values for both components and the ratio between the thermal flux and the total flux in Figure \ref{fig:flux}.

We note that the total soft X-ray flux initially increases until MJD 55154. The contribution of the \texttt{diskbb} component to the total flux also increases, which coincides with the spectral softening we note in Figure \ref{fig:HR}. For observations X2 and X4, where the thermal component is detected, the X-ray flux calculated for the \texttt{diskbb} component is around 26.2 to 53.2 percent of the total X-ray flux, respectively. 
After MJD 55154, the total flux remains somewhat constant until around MJD 55157. The flux then increases again, reaching a peak around MJD 55161, with XRT observations stopping after MJD 55164 due to Sun constraints. Using the \textit{RXTE} ASM count rate as presented in \cite{2010ApJ...716L.109M..MJ}, we confirm that the soft X-ray emission (2-10 keV) peaks around this time and decays afterwards. After MJD 55154, the \texttt{diskbb} component contributes considerably less to the total X-ray flux. The majority of XRT observations following MJD 55154 only have upper limits for the \texttt{diskbb} flux. For observations X13 to X15, where the thermal component is detected again, the \texttt{diskbb} component only contributes around 2.8 to 4.0 percent of the total X-ray flux. Therefore, the flux of the \texttt{nthcomp} component contributes substantially to the total flux overall. For most observations the flux for the \texttt{nthcomp} component is at least 91.6 to 99.7 percent of the total soft X-ray flux, considering the upper limit flux values for the \texttt{diskbb} component. 

\subsection{The radio/X-ray coupling}
\label{sec:results_lumplane}

\begin{figure}
 \includegraphics[width=\columnwidth]{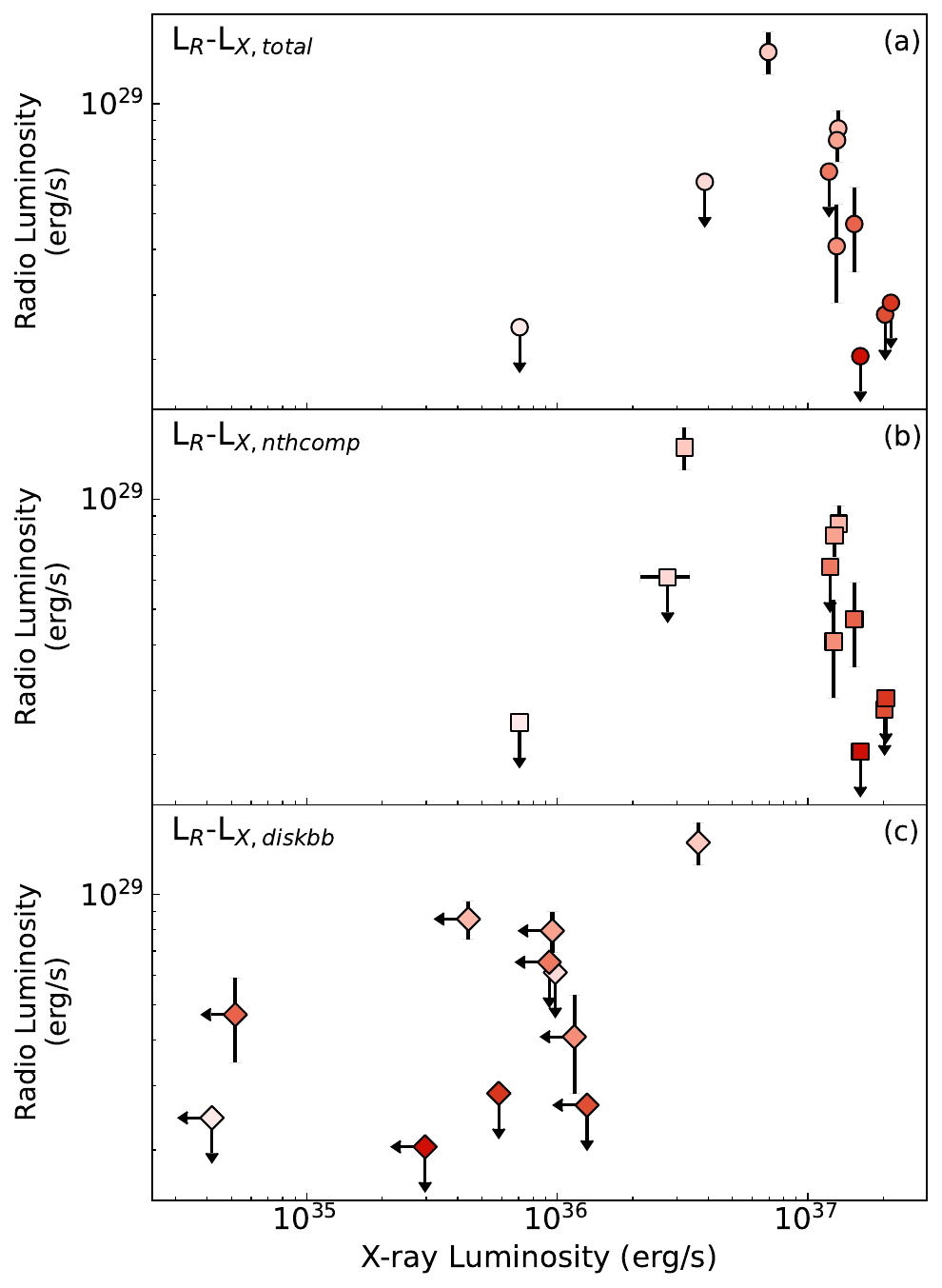}
 \caption{The coupling between the radio and X-ray luminosity, both expressed in erg/s, for the 2009 outburst of Aql X-1. The radio flux densities are obtained at 8.4 GHz and the X-ray flux is calculated between 0.7-10 keV. The evolution in time is indicated with red sequential colours from light to dark. (a) The coupling between the X-ray luminosity calculated for the complete composite model and the radio luminosity. (b) and (c) show this for the X-ray luminosity for the \texttt{nthcomp} and \texttt{diskbb} model components, respectively.} 
 \label{fig:lum_data}
\end{figure}

From our analysis, we find that fitting the composite model only intermittently provides a statistically improved fit to the Comptonised model. Moreover, we find that the flux of the \texttt{nthcomp} component contributes substantially to the total flux overall, and that the flux of the \texttt{diskbb} only contributes around 2.8 to 4.0 percent to the total flux, with the exception of observations X2 and X4.
Based on previous work, we find that this could be the result of degeneracies between the thermal and Comptonised component in the composite model, which we will discuss in more detail in Section \ref{sec:discussion}. 
As we aim to study if the presence of additional thermal emission affects the coupling of the radio/X-ray luminosity, we will explore what the radio/X-ray luminosity plane looks like for this work, keeping in mind that this could be affected by the aforementioned degeneracies.
We calculate the X-ray luminosity using the flux from the entire spectrum and from the \texttt{nthcomp} and \texttt{diskbb} components in the composite model separately. We show the resulting radio/X-ray luminosity plane in Figure \ref{fig:lum_data}.

From MJD 55143 to 55151 we see the total X-ray flux increase, including the flux in the \texttt{diskbb} component, as we described in the previous section. This is reflected in Figure \ref{fig:lum_data} as well for the first 2 data points, X1-R1 and X2-R4. During this period, we only measure upper limits for the radio emission. Then the first radio detection coincides with the XRT observation where the \texttt{nthcomp} and \texttt{diskbb} component contribute approximately equally to the total X-ray luminosity, this is data point X4-R5. This is also the observation where the \texttt{diskbb} component is the brightest, which clearly stands out in the Figure at $L_{\mathrm{R}} \approx 1.4 \times 10^{29} (d/ 4.5 \text{kpc})^2$ erg/s. During these observations, Aql X-1 is transitioning from a hard to a soft state. Once the source is in the soft state around MJD 55154, we note quenching in the radio emission as initially reported by \cite{2010ApJ...716L.109M..MJ}. Furthermore, we only have upper limits for the \texttt{diskbb} luminosity during this period, with the exception of X14-R14, around $L_{\mathrm{R}} \approx 2.9 \times 10^{28} (d/ 4.5 \text{kpc})^2$ erg/s. 

Overall, we find that with the exception of data point X4-R5, the \texttt{diskbb} luminosity does not contribute substantially to the total X-ray luminosity, nor trends we see in the radio/X-ray coupling before or after the spectral softening. We conclude that separately analysing the component probing the black body emission from the disc, neutron star surface and/or boundary layer, does not reveal a significantly changed radio/X-ray luminosity behaviour for this outburst of Aql X-1.

\section{Discussion}
\label{sec:discussion}

\subsection{Behaviour of the thermal component}
\label{sec:discussion_swift}

Using archival \textit{Swift}/XRT data, we have performed a detailed spectral analysis on the X-ray observations obtained from the 2009 outburst of Aql X-1. We have performed this research to study the relative contributions of the thermal black body emission and the Comptonised emission to the radio/X-ray connection derived for this outburst. We find that for these observations, fitting a composite (\texttt{tbabs*(diskbb+nthcomp)}) model intermittently provides a statistically improved fit to a Comptonised (\texttt{tbabs*nthcomp}) model.

As mentioned in Section \ref{sec:results}, the \textit{Swift}/XRT observations cover the rise and apparent peak of soft X-rays during the outburst. We initially observe a spectral softening around MJD 55150 to 55154, where the hard X-ray flux decays and the soft X-ray flux rises, as shown in Figure \ref{fig:lightcurves}. This softening is reflected in the soft (0.5-2.5 keV) and hard (2.5-10 keV) band of \textit{Swift}/XRT as shown in Figure \ref{fig:HR}. For two observations during this transition, namely observations X2 and X4, we find that the thermal black body component \texttt{diskbb} is a significant addition to the model, and the flux calculated for this component contribute arounds 26.2 and 53.2 percent to the total X-ray flux for X2 and X4, respectively. However, this is not the case for observation X3, which is performed during the spectral softening as well. Considering the upper limit found for the \texttt{diskbb} flux, the contribution is less than 37 percent for this observation.

For the majority of the observations following MJD 55154, we can only constrain upper limits for the \texttt{diskbb} component. For the observations where the \texttt{diskbb} component is a significant addition to the model, namely observations X13, X14 and X15, the flux calculated for the thermal component only contributes around 2.8 to 4.0 percent to the total X-ray flux. These observations occur around the peak total flux for the available XRT observations. For the remaining observations the contribution is less then 0.3 to 8.4 percent of the total X-ray flux, considering the upper limit values for the \texttt{diskbb} flux. We show the ratio between the \texttt{diskbb} and total X-ray flux in Figure \ref{fig:flux}c. Moreover, in Section \ref{sec:results_lumplane} we discuss the contribution of the \texttt{diskbb} luminosity, and conclude that separating the \texttt{nthcomp} and \texttt{diskbb} component does not reveal significantly changed radio/X-ray luminosity behaviour for this outburst.

To explore for the \textit{Swift}/XRT spectra if the thermal component is present but too faint to detect in the individual observations, we tried to combine the data of observations X5 to X8 and X9 to X12 and fit the composite and Comptonised model to the spectra. The $p$ value of fitting the composite model instead of the Comptonised model is 0.44 for the combined spectrum of observations X5 to X8, corresponding to a significance level of 0.15. The $p$ value of fitting the composite model to the combined spectrum of observations X9 to X12 is 1. This means that during observations X5 to X12, no significant thermal emission is present in the \textit{Swift}/XRT spectra. The upper limit of the flux in the \texttt{diskbb} component in these spectra is $f_{\rm th} < 1.16 \times 10^{-10}$ erg/s/cm$^2$ for the combined spectrum of observations X5 to X8 and $f_{\rm th} < 3.72 \times 10^{-11}$ erg/s/cm$^2$ for the combined spectrum of observations X9 to X12.

Based on these results, we find that the thermal component is rarely significant in fitting the XRT spectra, with the exception of a few observations. Furthermore, the flux of the thermal black body component contributes less than a few percent of the total X-ray flux in the soft X-ray band for most observations. We conclude that using a Comptonised model (\texttt{tbabs*nthcomp}) is mostly sufficient to fit \textit{Swift}/XRT spectra for outbursts of Aql X-1. We find no evidence of a significant thermal black body contribution in the \textit{Swift}/XRT spectra that could cause scatter in the radio/X-ray coupling of Aql X-1, neither in the soft nor in the hard state. 
 
However, we note that we would nominally expect a strong thermal component in the X-ray spectra recorded in the soft state. As spectral degeneracies have been identified between the thermal and Comptonised component in the past \citep[see e.g.][]{2007ApJ...667.1073L..Lin2007}, this could be affecting our results with regard to the thermal component.

\begin{figure}
 \includegraphics[width=0.95\columnwidth]{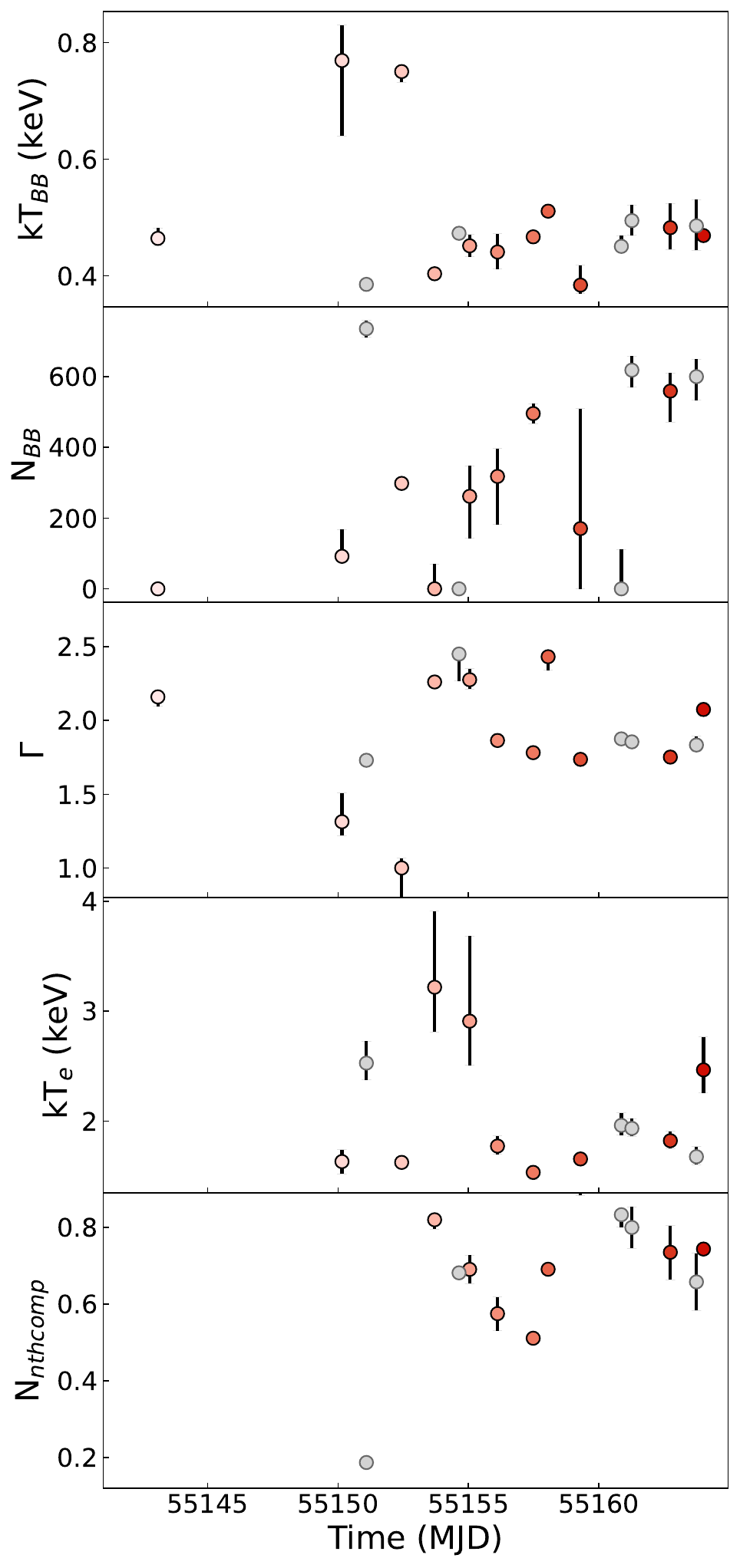}
 \caption{Parameter values of the composite model (\texttt{tbabs*(diskbb+nthcomp)}) for all \textit{Swift}/XRT spectra. 
  The \texttt{tbabs} model only has the parameter $n_{\mathrm{H}}$, which is fixed to $0.355 \times 10^{22}$ cm$^2$ for all calculations and fits. Parameters $kT_{\text{bb}}$ (in keV) and $N_{\text{BB}}$ correspond to the \texttt{diskbb} model component. Parameters \texttt{$\Gamma$}, $kT_{\text{e}}$ (in keV) and $N_{\text{nthcomp}}$ (unity at 1 keV for a norm of 1) correspond to the \texttt{nthcomp} model component.} 
 The colour coding of the data points used here is the same as in Figure \ref{fig:HR}.
 \label{fig:parameters}
\end{figure}

One way we can check for degeneracies, is by evaluating the parameter values from our fits. 
In Figure \ref{fig:parameters} we show the parameter values for the composite model. 
For some observations the value for $kT_{\text{e}}$ is pegged at the maximum value for the parameter, around 999 keV. We expect that for these observations no reliable value for the $kT_{\text{e}}$ parameter could be found, possibly because the electron temperature exceeds 10 keV, i.e. outside the XRT passband. Therefore, these data points are not included in the $kT_{\text{e}}$ panel and are set to \texttt{NaN}. 
For one of the observations, X16, the $N_{\text{BB}}$ parameter is also set to \texttt{NaN}, as adding a \texttt{diskbb} component to the model resulted in a $p$ value of 1. Because the model component is not needed to fit the data properly, this appears to cause some degeneracy in the $N_{\text{BB}}$ parameter for this observation. Furthermore, we also note significant changes in parameter values between consecutive observations. For example, we see the $kT_{\text{bb}}$ and $N_{\text{BB}}$ parameter change significantly from observations X2 to X3 and X3 to X4. As we mentioned in Section \ref{sec:datamethod}, this is a known parameter degeneracy, which does not affect the calculations of the flux of the model component. Lastly, we find some degeneracy between the $kT_{\text{bb}}$ and $\Gamma$ parameters for observations X2 to X4 as well.

Based on these findings, we note degeneracies for some fits for the parameters of the \texttt{diskbb} and \texttt{nthcomp} component. However, these might also be indicative of degeneracies between the thermal and Comptonised component, as we note in Section \ref{sec:results_model}. 
As the energy range in the \textit{Swift}/XRT spectra is limited between 0.7-10 keV, this might affect our ability to constrain the significance of adding a thermal component to the Comptonised model.

To this end, it is important to consider that for our analysis of \textit{Swift}/XRT, when considering a standard significance level of 3$\sigma$ to add an additional \texttt{diskbb} component to the Comptonised model, only observations X4 and X13 satisfy this requirement. For this analysis, we have considered X2, X14 and X15 to have a significant thermal black body component as well, while the significance level of adding this component ranges between 2.5$\sigma$ and 3$\sigma$. We chose to include these observations as we do not find a comparable significance level for the other observations. In this work we note, for example, that the upper limits we find are not always constraining. We show in Figure \ref{fig:lum_data}c that for some observations, the calculated value for the \texttt{diskbb} flux is lower than the upper limits determined for other observations. Consequently, we cannot establish if additional thermal emission is indeed not significant in the latter observations.

To test if adding higher energy coverage might resolve any degeneracies between the thermal and Comptonised component in the composite model, we perform joint fits with quasi-simultaneous \textit{RXTE} data, which are also presented in \citep{2010ApJ...716L.109M..MJ}.

\subsection{Joint fits \textit{Swift} and \textit{RXTE} spectra}
\label{sec:discussion_jointfits}
The \textit{RXTE}/PCA observations are described in Section \ref{sec:datamethod}. 
We matched \textit{RXTE}/PCA and \textit{Swift}/XRT observations if they were measured $\la 1$ day from each other. The resulting matched observations are listed in Table \ref{tab:RXTEobs}, and we indicate these with lowercase x1 to x16. To perform the joint fits, we used the same methods as described in Section \ref{sec:datamethod}. We added a constant factor to both the \textit{RXTE} and \textit{Swift} spectra, and froze this factor for the \textit{Swift} data at 1, to account for differences in sensitivity and calibration while fitting. Moreover, we also included a \texttt{gauss} model component to the model for the \textit{RXTE} data, to account for an Fe-K reflection emission line around 6.7 keV \citep[see e.g.][]{2007ApJ...667.1073L..Lin2007}. 
We did not fit a gaussian to the \textit{Swift} data, as the sensitivity around 6.7 keV is lower  than for \textit{RXTE}, and no gaussian was needed to fit the \textit{Swift} data properly. The \textit{Swift}/XRT spectra are fit over the range 0.7-10 keV, and the \textit{RXTE} PCA spectra over the range 2.6-23 keV. Finally, we calculated the flux over the 0.7-10 keV range, to allow comparison of the joint fits to the results of only fitting the \textit{Swift}/XRT data. 

\begin{table}
\resizebox{\columnwidth}{!}{%
\begin{tabular}{@{}clccc@{}}
\toprule
                         & X-ray & MJD RXTE & Time difference   & MJD Swift \\
                         & obs   & (day)    & Swift-RXTE (hour) & (day)     \\ \midrule
\cellcolor[HTML]{FAEFFF} & x1    & 55143.07 & 0.51              & 55143.09  \\
\cellcolor[HTML]{F1E0F7} & x2    & 55150.28 & 3.15              & 55150.15  \\
                         & x3$^{a}$    & 55151.19 & 2.51              & 55151.09  \\
\cellcolor[HTML]{DFC2E8} & x4    & 55152.17 & 6.42              & 55152.44  \\
\cellcolor[HTML]{C79CD5} & x5    & 55153.08 & 14.9              & 55153.70  \\
                         & x6$^{a}$    & 55154.6  & 0.89              & 55154.64  \\
\cellcolor[HTML]{B47FC6} & x7    & 55155.04 & 0.11              & 55155.04  \\
\cellcolor[HTML]{AB70BF} & x8    & 55156.09 & 0.53              & 55156.11  \\
\cellcolor[HTML]{A264B9} & x9    & 55157.14 & 8.18              & 55157.48  \\
\cellcolor[HTML]{9A58B3} & x10   & 55157.85 & 4.86              & 55158.05  \\
\cellcolor[HTML]{924CAC} & x11   & 55158.97 & 7.78              & 55159.29  \\
                         & x12$^{a}$   & 55160.15 & 17.1              & 55160.86  \\
                         & x13$^{a}$   & 55161.07 & 4.66              & 55161.26  \\
\cellcolor[HTML]{883DA5} & x14   & 55163.03 & 6.99              & 55162.74  \\
                         & x15$^{a}$   & -        & -                 & 55163.74  \\
\cellcolor[HTML]{7E2C9E} & x16   & 55163.94 & 1.67              & 55164.01 
\end{tabular}%
}
\caption{Details of the \textit{RXTE} observations and the corresponding \textit{Swift} observations. The MJD are provided for the middle of the X-ray observations. 
The X-ray observations measured $\la 1$ day apart are matched for the joint fits. These observations are indicated with a lowercase x, in contrast to the uppercase X indicated in Table \ref{tab:observations}. The data that can be matched with quasi-simultaneous radio data, as described in Table \ref{tab:observations}, are indicated with purple sequential colours from light to dark to indicate the evolution in time in later Figures.
\\ \hspace{\textwidth} $^{a}$  These X-ray observations could not be matched to a Radio observation and are therefore excluded from this analysis.}
\label{tab:RXTEobs}
\end{table}

We find that fitting the composite model provides a statistically improved fit to the Comptonised model for more observations than we find in our previous results, as shown in Table \ref{tab:RXTE_sign}. The thermal component is  significant for the joint fits to observations x2-x4, x8-x9 and x11-x16. Only for observations x1, x5-x7, and x10 is the Comptonised model statistically preferred. We note this in particular for observations x5 to x7, as these observations directly follow the transition from the intermediate to the soft state, as discussed in Section \ref{sec:results_model}.

\begin{table}
\resizebox{\columnwidth}{!}{%
\begin{tabular}{@{}lcccc@{}}
\toprule
Obs & $\chi_{2}$ dual & $\chi_{2}$ single & \multicolumn{1}{l}{p-value} & Significance level \\
    & model (d.o.f.) & model (d.o.f.)   & \multicolumn{1}{l}{}        & p value ($\sigma$) \\ \midrule
x1  & 1100.4 (772)   & 1100.4 (773)     & 1                           & -                  \\
x2  & 959.43 (850)   & 1047.37 (851)    & 6.06E-18                    & 8.55               \\
x3$^{a}$  & 710.59 (609)   & 835.04 (610)     & 3.78E-23                    & 9.84               \\
x4  & 1117.11 (907)  & 1137.29 (908)    & 5.61E-05                    & 3.86               \\
x5  & 1103.4 (896)   & 1103.4 (897)     & 1                           & -                  \\
x6$^{a}$  & 1171.13 (924)  & 1171.13 (925)    & 1                           & -                  \\
x7  & 907.46 (917)   & 914.24 (918)     & 9.00E-03                    & 2.37               \\
x8  & 951.79 (883)   & 1071.56 (884)    & 1.49E-24                    & 10.2               \\
x9  & 923.12 (849)   & 992.27 (850)     & 4.92E-15                    & 7.74               \\
x10 & 1533.6 (939)   & 1533.6 (940)     & 1                           & -                  \\
x11 & 1013.12 (892)  & 1140.99 (893)    & 7.50E-25                    & 10.2               \\
x12$^{a}$ & 941.97 (899)   & 1007.55 (900)    & 7.46E-15                    & 7.69               \\
x13$^{a}$ & 1111.92 (931)  & 1578.07 (932)    & 7.95E-73                    & 18.0               \\
x14 & 1055.74 (921)  & 1433.79 (922)    & 3.12E-63                    & 16.7               \\
x15$^{a}$ & -              & -                & \multicolumn{1}{l}{-}       & -                  \\
x16 & 938.38 (906)   & 952.18 (907)     & 2.77E-04                    & 3.45               \\ \bottomrule
\end{tabular}%
}
\caption{Details of the joint X-ray spectral fits to the \textit{Swift}/XRT and \textit{RXTE}/PCA spectra for the composite (\texttt{tbabs*(nthcomp+diskbb)}) and Comptonised model (\texttt{tbabs*nthcomp}). The probability of the F-statistic, or \textit{p} value, is obtained by comparing the $\chi^{2}$ values and degrees of freedom (d.o.f.) obtained for both models for each observation using \texttt{ftest} in \texttt{XSPEC}. The significance level is obtained using a standard normal distribution ($\mu$=0, $\sigma$=1) for the corresponding probability with a one-tailed test.
\\\hspace{\textwidth} $^{a}$  These X-ray observations could not be matched to a Radio observation and are therefore excluded from this analysis.}
\label{tab:RXTE_sign}
\end{table}

\begin{figure}
 \includegraphics[width=\columnwidth]{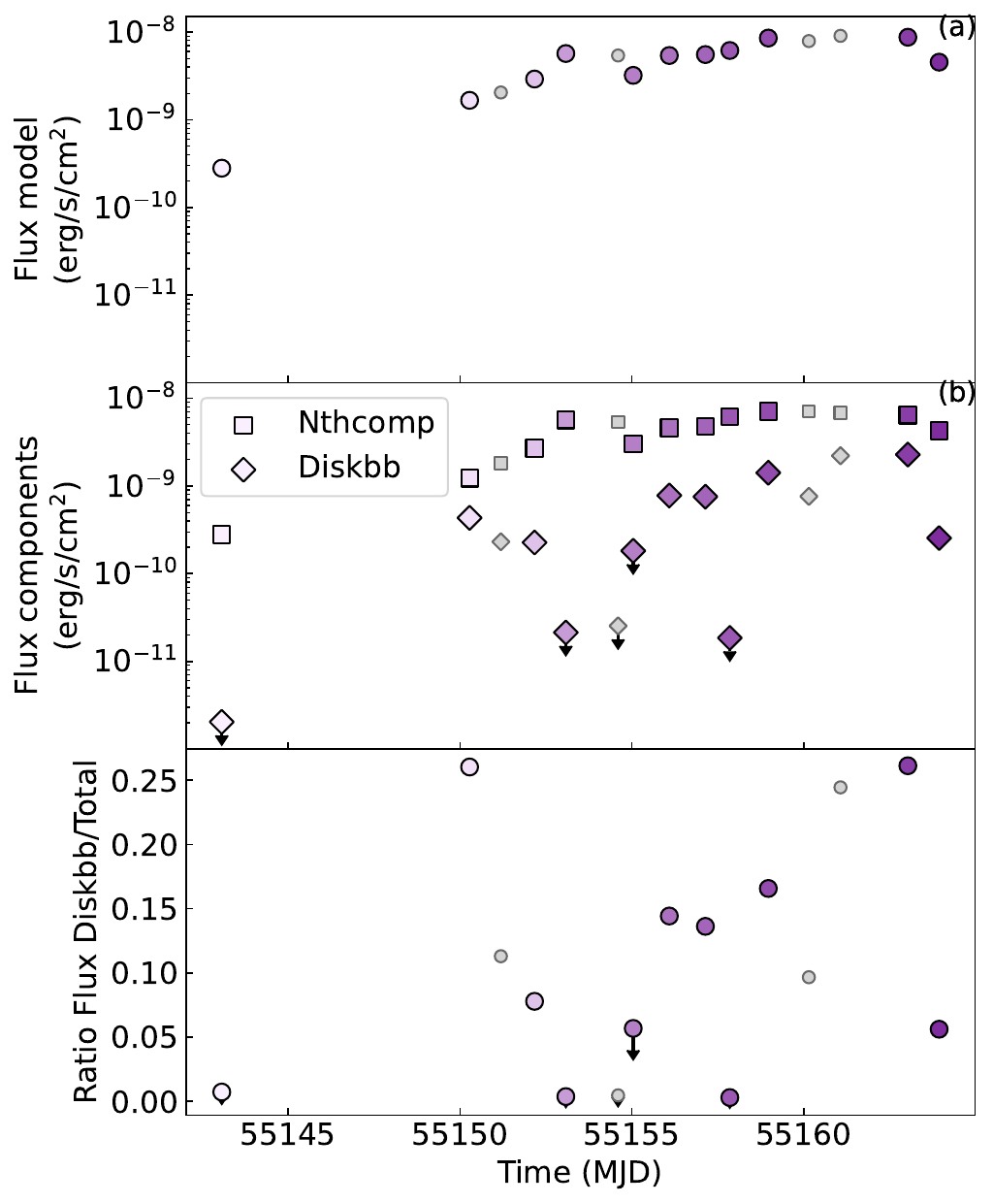}
 \caption{The X-ray flux values obtained for the joint fits to the \textit{Swift}/XRT and \textit{RXTE}/PCA spectra for composite model and the separate model components between 0.7-10 keV. 
 The evolution in time is indicated with purple sequential colours from light to dark. The large coloured points indicate the X-ray observations that could be linked to quasi-simultaneous radio observations. Small grey points indicate X-ray observations that could not be linked. (a) The X-ray flux calculated for the complete composite model from 0.7-10 keV. (b) The X-ray flux for the \texttt{diskbb} and \texttt{nthcomp} model components. The \texttt{diskbb} and \texttt{nthcomp} component are indicated using diamonds and squares, respectively. (c) The ratio between the X-ray flux for the \texttt{diskbb} component and the total flux.}  
 \label{fig:flux_jointfit}
\end{figure}

\begin{figure}
 \includegraphics[width=\columnwidth]{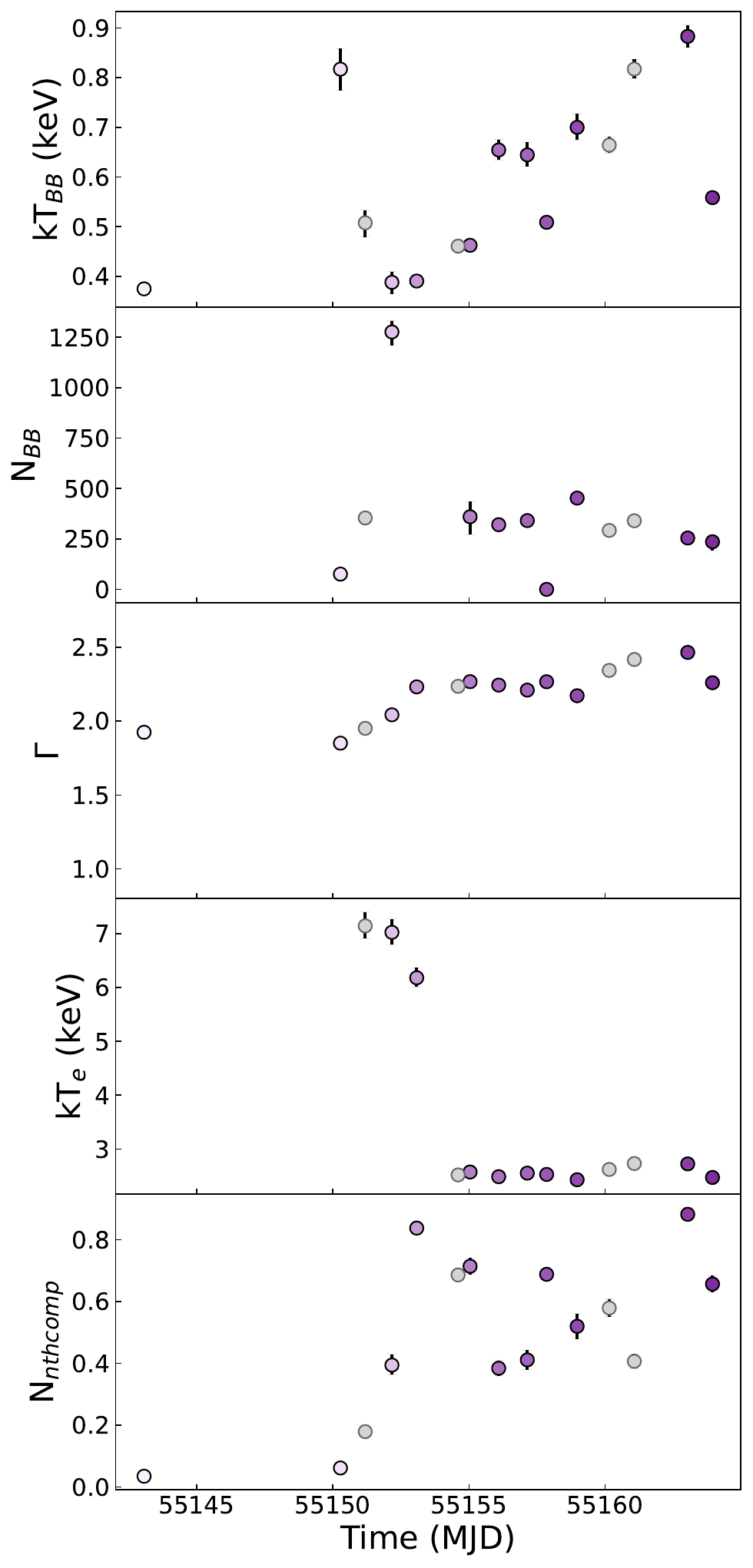}
 \caption{Parameter values obtained for the joint fits to the \textit{Swift}/XRT and \textit{RXTE}/PCA spectra for composite model. The \texttt{tbabs} model only has the parameter $n_{\mathrm{H}}$, which is fixed to $0.355 \times 10^{22}$ cm$^2$ for all calculations and fits. Parameters $kT_{\text{bb}}$ (in keV) and $N_{\text{BB}}$ correspond to the \texttt{diskbb} model component. Parameters \texttt{$\Gamma$}, $kT_{\text{e}}$ (in keV) and $N_{\text{nthcomp}}$ (unity at 1 keV for a norm of 1) correspond to the \texttt{nthcomp} model component. The colour coding of the data points used here is the same as in Figure \ref{fig:flux_jointfit}.}
 \label{fig:parameters_jointfit}
\end{figure}

We show the flux for the total model and the individual components in Figure \ref{fig:flux_jointfit}, and the parameter values for the composite model in Figure \ref{fig:parameters_jointfit}. We note that the flux of the \texttt{diskbb} component increases in the soft state, but that at the start of the soft state (x5-x7) we only have upper limits for the thermal flux. We note that these fits can still be affected by spectral degeneracies between the thermal and Comptonised component as discussed previously. However, we also note that the upper limits found for the \texttt{diskbb} flux are more constraining than those for the \textit{Swift}/XRT spectra. We also find that the ratio between the \texttt{diskbb} flux and the total flux is around $26\%$ at x2, after which the contribution decreases, and gradually increases again in the soft state, contributing around $26.1\%$ at the peak of the soft X-ray flux around x14. 

\begin{figure}
 \includegraphics[width=\columnwidth]{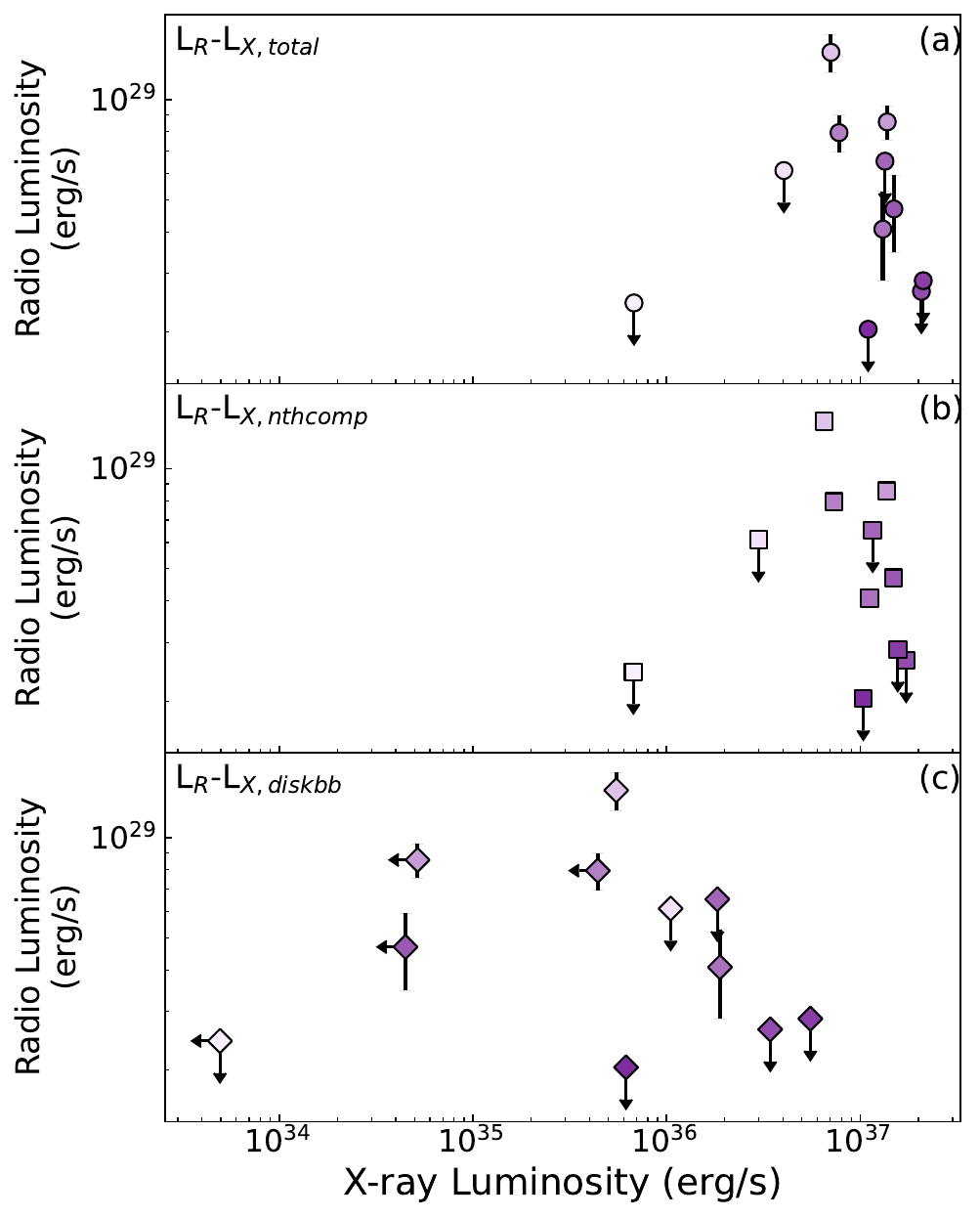}
 \caption{The coupling between the radio and X-ray luminosity, where the X-ray luminosity is obtained from the joint fits to the \textit{Swift}/XRT and \textit{RXTE}/PCA spectra. The radio flux densities are obtained at 8.4 GHz and the X-ray flux is calculated between 0.7-10 keV. The evolution in time is indicated with purple sequential colours from light to dark. (a) The coupling between the X-ray luminosity calculated for the complete composite model and the radio luminosity. (b) and (c) show this for the X-ray luminosity for the \texttt{nthcomp} and \texttt{diskbb} model components, respectively.} 
 \label{fig:lum_data_jointfit}
\end{figure}

Putting the results of the joint fits in the context of the radio/X-ray luminosity plane in Figure \ref{fig:lum_data_jointfit}, we see that the luminosity for the \texttt{diskbb} component increases as the radio luminosity quenches. However, we still find that the radio/X-ray coupling is not significantly changed when we separate emission from a thermal component.

Overall, performing joint fits with \textit{Swift}/XRT and \textit{RXTE}/PCA appears to have resolved some of the spectral degeneracies that might have affected the fits to the \textit{Swift}/XRT spectra. However, our conclusions regarding the radio/X-ray coupling remain unchanged. Moreover, we still find that the thermal component is not always a significant addition to modelling the X-ray spectra. Therefore, this analysis should be followed up with more sensitive, broad-band X-ray observations and densely-sampled near-simultaneous radio observations in the future (see Section~\ref{sec:discussion_future_research}).
\\

\subsection{Interpretation jet/accretion behaviour}
\label{sec:discussion_jet-accretion_behaviour}

Since our results can be affected by degeneracies between the thermal and Comptonised component, there are issues in constraining the contribution of the thermal component in our X-ray spectra. Therefore, it is difficult to interpret the results to study the physical processes. However, we can consider the implications of our results to the radio/X-ray coupling, to discuss what this would mean for the jet/accretion behaviour if these degeneracies were not present.

In this case, our results seem to partially agree with the behaviour of the accretion flow and jets in some BH-LMXB sources. For transient outbursts of BH-LMXBs, one observes compact jets during the hard state, and the jets persist until the X-ray spectrum softens \citep{2004MNRAS.355.1105F..HID}. Once the source transitions to the soft state, one notes quenching of the jet 
\citep[see e.g.][]{2003MNRAS.344...60G, 2011ApJ...739L..19R..quenchingdiscussion, 2011MNRAS.414..677C..quenchingdiscussion}.
For some sources, radio emission has been detected with a negative spectral slope during the bright hard to soft state transition near the peak of the outburst. This is interpreted as being associated with the scenario where the core jet is suppressed once the accretion disc propagates inwards toward the innermost stable circular orbit (ISCO). In this scenario the optically thin radio emission is associated with jet ejecta probing the re-brightening at shocks from the jet propagating away from the BH-LMXB \citep{2004MNRAS.355.1105F..HID, 2009MNRAS.396.1370F...ejectaexplained}. 
When the source transitions back to the hard state, one observes compact jets again \citep[see e.g.][]{2005ApJ...622..508K..BHjetreturns, 2013ApJ...779...95K..BHjetreturns}.

For Aql X-1, some of these features have been observed as well. \cite{2018A&A...616A..23D..quench} reports radio to millimetre emission with a negative slope during the 2016 outburst of Aql X-1 after the transition to the soft state. Furthermore, \cite{2009MNRAS.400.2111T..critX-raylum} has studied the outbursts in 2002, 2004 and 2005 using multiwavelength data, and found that the radio emission quenched above a critical X-ray luminosity of around 10 percent of the Eddington luminosity. \cite{2009MNRAS.400.2111T..critX-raylum} suggest this could be similar behaviour as seen for BH-LMXBs like we mentioned previously. 

For the 2009 outburst, only one constraint could be placed on the spectral slope of the radio emission. Using the 8.4 GHz VLBA observation (labelled R7 in this work) and the 5 GHz EVN observation between MJD 55154 and 55155, \cite{2010ApJ...716L.109M..MJ} find a slightly inverted slope of $\alpha = 0.40 \pm 0.31$, which is detected after the transition to the soft state.

We are particularly interested in the radio observations during the bright transition from the hard to the soft state, to see if Aql X-1 could show transient jet ejecta like those observed for BH-LMXBs. Unfortunately, we have no constraints on the spectral slope of the radio emission in this period. However, we do find that the first and brightest detection of radio emission R5 coincides with X4/x4. Based on our spectral analysis of the \textit{Swift}/XRT spectra, we find that X4 has the brightest thermal component. From our joint fits with the \textit{RXTE} data, we find a significantly lower \texttt{diskbb} flux for x4. However, we also see that between X1 and X4 the soft colour for \textit{Swift}/XRT increases, and peaks for observation X4. This could be indicative of the disc moving toward the ISCO during the transition from the hard and the soft state, i.e. between X1 and X4, causing the material in the disc to heat up and the thermal component to become brighter. This could potentially result in more flux in the soft X-ray band (0.5-2.5 keV) when compared to the hard X-ray band (2.5-10 keV). 

Following radio observation R5, we note quenching in the radio emission, as initially reported by \cite{2010ApJ...716L.109M..MJ}. This happens at a total X-ray luminosity of around $L_{\mathrm{X}} = 1.3 \times 10^{37} (d/ 4.5 \text{kpc})^2$ erg/s. This also fits the scenario seen for BH-LMXBs, where the radio emission quenches once the source transitions to the soft state. Once the source is in the soft state and the jet quenches, we mainly find upper limits for the luminosity of the \texttt{diskbb} component, with the exception of observations X13 to X15, when we only look at the results for the \textit{Swift}/XRT spectra. For the joint fits with \textit{Swift}/XRT and \textit{RXTE}/PCA, we find that $kT_{\text{bb}}$ and the contribution of the thermal component increases in the soft state from x7 on. In both cases, the detected thermal emission could probe the disc heating up again, as it coincides with the apparent peak of the soft X-ray emission. 

We also consider if the thermal emission in observations X2 to X4 probes another emission process than the thermal emission in later observations, as we only find a significant thermal component in X13 to X15 for the \textit{Swift}/XRT spectra. For example, \citet{2015MNRAS.454.1371W} found that for the accretion flow of NS-LMXBs at low luminosities, the thermal emission from the neutron star surface can dominate the spectrum. To explore this scenario, we may study the parameter values obtained for the composite model, especially the $kT_{\text{bb}}$ parameter. However, due to parameter degeneracies as discussed in Section \ref{sec:discussion_swift}, we were unable to determine this for the \textit{Swift}/XRT spectra. 
For the joint \textit{Swift}/XRT and \textit{RXTE}/PCA fits, we find that $kT_{\text{bb}}$ is around 0.77 keV at x2, then the temperature appears to decrease, and gradually increases again in the soft state. A similar pattern in seen for the contribution of the \texttt{diskbb} flux to the total flux. However, this could again be the result of the parameter degeneracy between $kT_{\text{bb}}$ and $N_{\text{BB}}$. For x4, we note a relatively low value for $kT_{\text{bb}}$, and a peak value for the $N_{\text{BB}}$ parameter. Therefore, we are unable to link these results for the thermal emission to a specific emission process.

Based on our results, we cannot confirm that the thermal emission detected in the \textit{Swift}/XRT spectra nor the joint \textit{Swift}/XRT and \textit{RXTE}/PCA fits probe the accretion disc in particular.  
The scenario proposed for the jet/accretion coupling for some BH-LMXBs appears to match our results, where the disc moves toward the ISCO and bright transient jet ejecta are observed during the transition from the hard to the soft state, and later the radio emission from the jet quenches in the soft state. 
However, we need more information to unambiguously determine the jet/accretion behaviour for Aql X-1, and to be able to compare the behaviour of BH-LMXB and NS-LMXB sources.

\subsection{Future research}
\label{sec:discussion_future_research}

\noindent More sensitive observations could allow us to identify separate spectral components probing the emission of the NS surface, boundary layer and accretion disc, while only one thermal component could be identified in this work. One telescope that we propose for follow-up research is the \textit{Neutron star Interior Composition Explorer} \cite[\textit{NICER}, ][]{2016SPIE.9905E..1HG...NICER}. Simple spectral simulations, detailed in Appendix \ref{app:NICER}, show how short \textit{NICER} monitoring observations are able to establish the presence of a thermal component down to significantly lower thermal luminosities. An extension of this work using \textit{NICER} and radio monitoring therefore allows for more detailed follow-up studies. However, we also stress that these simulations operate under the assumption that the input model is correct, i.e. that only statistical uncertainties play a role. That assumption implies that the complications with this analysis, such as model degeneracies in the absence of hard X-ray coverage, will remain for NICER as well.

These issues of parameter degeneracy and broad-band energy converage may be alleviated more effectively through future missions. For instance, the proposed \textit{Strobe-X} mission \citep{strobex} is planned to combine a soft (0.2-12 keV) instrument, the X-ray Concentrator Array, as well as a harder (2-30 keV) instrument, the Large Area Detector. Alternatively, the proposed \textit{HEX-P} mission is planned to combine two detectors, the Low Energy Telescope and the High Energy Telescope, to provide simultaneous spectral coverage between 0.1 and 150 keV \citep{hexp}. The broad-energy coverage of both observatories may remove the issues found in our work with model degeneracies, while syngergies with next generation radio observatories could allow for follow up studies at unprecedented sensitivity. An essential requirement, for this purpose, however is sufficient flexibility to allow for high cadence monitoring of an LMXB outburst, similar to \textit{RXTE}, \textit{Swift}, and \textit{NICER}.

\section*{Data availability}
The data underlying this article will be available in Zenodo at DOI: 10.5281/zenodo.6477045 upon publication. These astrophysical data sets were derived from archival radio data reported by \cite{2010ApJ...716L.109M..MJ} and X-ray data from sources in the public domain: \url{https://heasarc.gsfc.nasa.gov/cgi-bin/W3Browse/swift.pl}. 

\section*{Acknowledgements}
We thank the anonymous referee for the valuable comments.
SF and ND acknowledge the hospitality of the University of Oxford, where part of this research was carried out, and support from the Leids Kerkhoven-Bosscha Fonds. JvdE and ND were supported by a Vidi grant from the Netherlands Organization for Scientific Research (NWO) awarded to ND when part of this research was carried out. JvdE is also supported by a Lee Hysan Junior Research Fellowship awarded by St. Hilda's College, Oxford. TDR acknowledges financial contribution from the agreement ASI-INAF n.2017-14-H.0.
This work made use of data supplied by the UK Swift Science Data Centre at the University of Leicester.

This is a pre-copyedited, author-produced PDF of an article accepted for publication in MNRAS following peer review. The version of record Fijma et al., 2023, MNRAS, 521, 4490 is available online at: \url{https://academic.oup.com/mnras/article/521/3/4490/7055950}.









\appendix

\section{\textit{NICER} simulations}
\renewcommand{\thefigure}{A\arabic{figure}}
\label{app:NICER}
\setcounter{figure}{0}

\begin{figure*}
 \begin{center}
	\includegraphics[width=8.5cm]{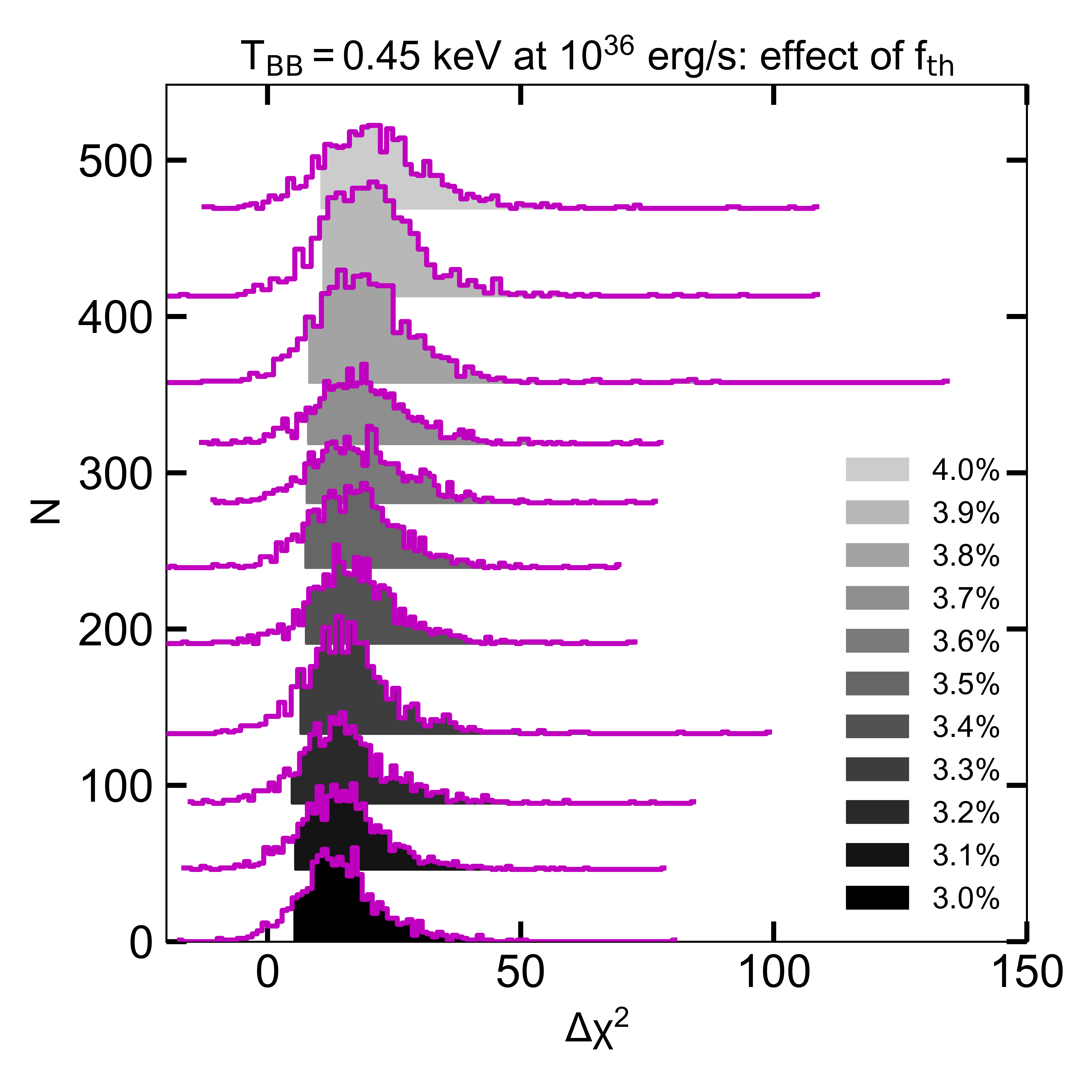}\hspace{+0.0cm}
		\includegraphics[width=8.5cm]{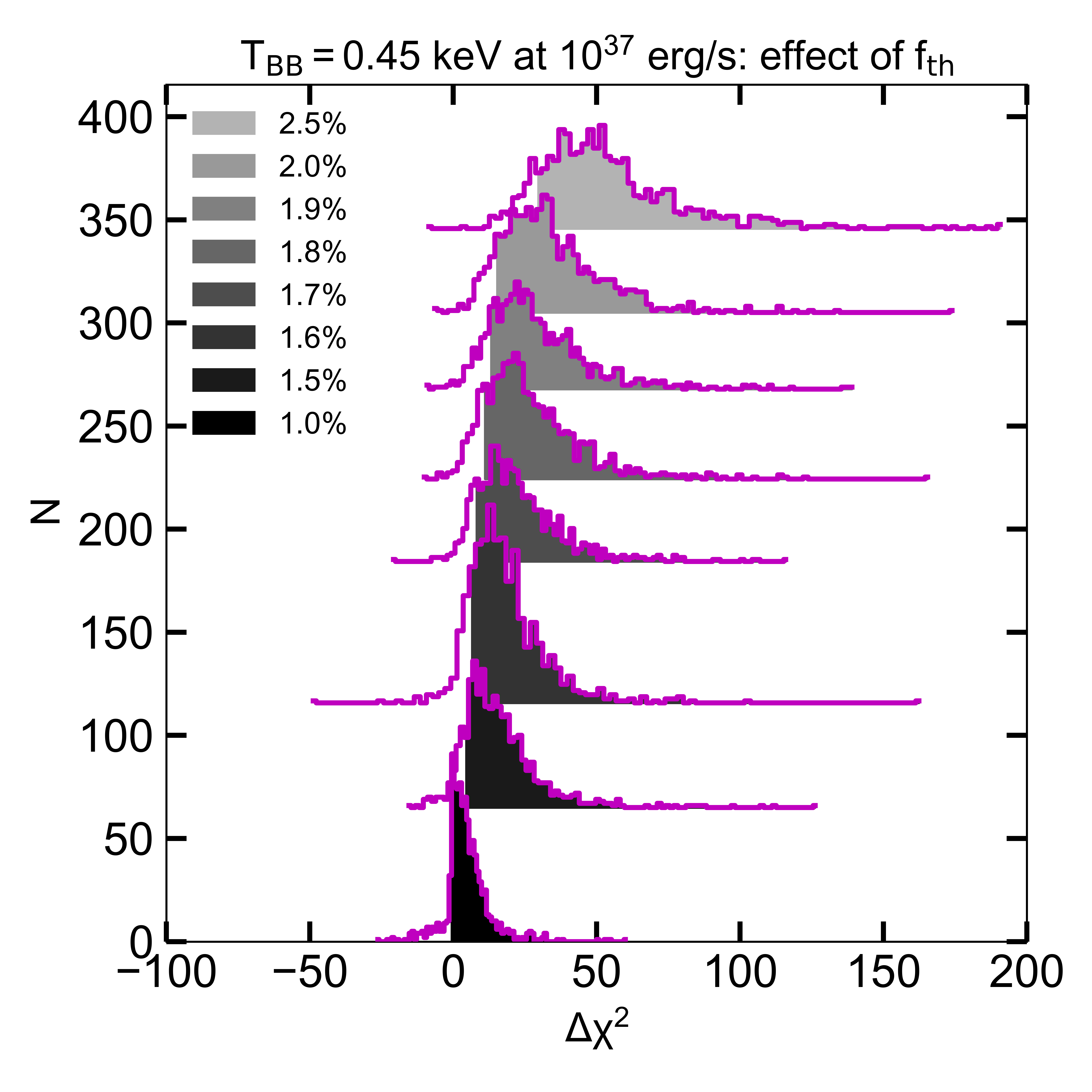}
    \end{center}
    \caption[]{Histograms of the fit improvement $\Delta \chi^2$ when comparing fits of a synthetic \textit{NICER} spectrum, simulated with the indicated $L_{\mathrm{X}}$ and $f_{\rm th}$, with both an absorbed \texttt{nthcomp} and the composite model. The histograms are vertically offset for clarity. The shaded region includes the highest $90$\% of $\Delta \chi^2$ values.}
 \label{fig:simulations}
\end{figure*}

In order to assess whether the \textit{Neutron star Interior Composition Explorer} \cite[\textit{NICER}, ][]{2016SPIE.9905E..1HG...NICER} could be used to established the presence of thermal emission to lower X-ray fluxes, we performed a suite of X-ray spectral simulations. Using the \texttt{fakeit} task in \texttt{xspec}, we simulated spectra following the \texttt{tbabs*(diskbb+nthcomp)} model used in this work. Each simulated spectrum was then fitted with both the full model, assuming the same constraints as in the main paper (i.e. fixed $n_{\mathrm{H}} = 0.355 \times 10^{22}$ cm$^{-2}$, linked disc and seed photon temperatures), and the simpler \texttt{tbabs*nthcomp} model. Linking the temperatures means that the composite model has one more free parameter, implying that a fit improvement of $\Delta \chi^2 = 9$ is significant at the $3\sigma$ level. With our simulations, we set out to find the minimum fractional flux in the thermal component, that would leave the thermal component $3\sigma$ detectable in at least $90\%$ of the simulated spectra. 

As input for the simulated spectra, we assumed either $L_{\mathrm{X}} = 10^{37}$ erg/s or $L_{\mathrm{X}} = 10^{36}$ erg/s. We set $\Gamma = 2$, $kT_e = 1.8$ keV, and $kT_{\rm BB} = 0.45$ keV, as typical values found in our analysis of Aql X-1. We also assumed $d = 4.5$ kpc to convert luminosities to fluxes. The normalisations of the \texttt{diskbb} and \texttt{nthcomp} components were determined such that the thermal, unabsorbed flux equaled $f_{\rm th} \times L_X / 4\pi d^2$, compared to a non-thermal unabsorbed flux of $(1-f_{\rm th})\times L_{\mathrm{X}} / 4\pi d^2$. In the simulations, we used the \textit{NICER} response and background files for an observation of Aql X-1 with an exposure of $\sim 1$ ks (ObsID 2050340127): after accessing the observation files, we ran \texttt{nicerl2} to re-reduce the data and subsequently applied standard routines using \texttt{nicerarf}, \texttt{nicerrmf}, and the \texttt{python} package \texttt{nicergof.bkg} to generate the arf and rmf files, and background spectrum. Each spectrum was similated for an exposure time of $1$ ks; for each setup of input parameters ($L_{\mathrm{X}}$, $f_{\rm th}$) we performed 1000 runs, each run saving the fit improvement $\Delta \chi^2$. 

In Figure \ref{fig:simulations}, we show the histograms of the resulting $\Delta \chi^2$ values for each combination of $L_{\mathrm{X}}$ ($10^{36}$ erg/s left; $10^{37}$ erg/s right) and $f_{\rm th}$. The histograms are vertically offset for clarity. The shaded regions, corresponding to the $f_{\rm th}$ indicated in the legend, indicate $90$\% of the $1000$ runs. As we define $\Delta \chi^2 \geq 9$ as a significant detection of the thermal component, we search for the minimum $f_{\rm th}$ where the shaded region lies fully at $\Delta \chi^2 \geq 9$. We find that at $L_{\mathrm{X}} = 10^{36}$ erg/s, $f_{\rm th} \geq 3.9$\% is detectable in $\geq 90$\% of simulated spectra. For $L_{\mathrm{X}} = 10^{37}$ erg/s, this limit drops to $f_{\rm th} \geq 1.8\%$. These limits are significantly lower than most of the \textit{Swift} limits reported in the main paper (c.f. Figure \ref{fig:lightcurves}, bottom panel; $L_{\mathrm{X}}=10^{37}$ erg/s corresponds to $\sim 4 \times 10^{-9}$ erg/s/cm$^2$), especially at the higher of the two luminosities. 

These results are simplified, in terms of the other spectral parameters: the exact values of the other broadband parameters will influence the actual significance of a thermal component, especially close to the $3\sigma$ threshold. However, these simulations indicate the promise of using \textit{NICER} monitoring of an extension of the main paper to other sources and outbursts.


\bsp	
\label{lastpage}
\end{document}